\documentclass[10pt,journal]{IEEEtran}

\usepackage{bm} 
\usepackage{xspace}

\newcommand{\xmath}[1] {\ensuremath{#1}\xspace}
\newcommand{\blmath}[1] {\xmath{\bm{#1}}}

\newcommand{\fb}{{\blmath f}}

\newcommand{\lb}{{\blmath l}}

\newcommand{\xb}{{\blmath x}}

\newcommand{\Rd}{{\mathbb R}}

\newcommand{\Qc}{{{\mathcal Q}}}

\newcommand{\beq}{\begin{equation}}
\newcommand{\eeq}{\end{equation}}
\newcommand{\beqa}{\begin{eqnarray}}
\newcommand{\eeqa}{\end{eqnarray}}

\usepackage{cite}
\usepackage{amsmath,amssymb,amsfonts}
\usepackage{mathrsfs}
\usepackage{algorithmic}
\usepackage{graphicx}
\usepackage{textcomp}
\usepackage{multirow}
\usepackage{multicol}
\usepackage{dsfont}
\usepackage{array} 
\usepackage{booktabs}
\usepackage{subcaption}
\usepackage{float}
\usepackage[final]{changes}
\usepackage{marginnote}
\usepackage{color}

\def\BibTeX{{\rm B\kern-.05em{\sc i\kern-.025em b}\kern-.08emT\kern-.1667em\lower.7ex\hbox{E}\kern-.125emX}}

\usepackage{cite}
\usepackage{bm} 
\usepackage{hyperref}

\def \paramone {Lung areas intensity}
\def \paramtwo {Lung areas intensity variance}
\def \paramthree {Cardiothoracic ratio}

\begin{document}

\title{Deep Learning COVID-19 Features
on CXR using Limited Training Data Sets}

\author{Yujin Oh$^1$,  Sangjoon Park$^1$, and Jong Chul Ye, \IEEEmembership{Fellow, IEEE}
\thanks{$^1$: Co-first authors with equal contribution. 
Authors are with the Department of Bio and Brain Engineering, Korea Advanced Institute of Science and Technology (KAIST), Daejeon 34141, Republic of Korea (E-mail: \{yujin.oh,depecher,jong.ye\}@kaist.ac.kr).}}
\maketitle

\begin{abstract}
Under the global pandemic of COVID-19, the use of artificial intelligence to analyze chest X-ray (CXR) image for COVID-19 diagnosis and patient triage is becoming important. Unfortunately, due to the emergent nature of the COVID-19 pandemic, a systematic collection of  CXR data set for deep neural network training is difficult. To address this problem, 
here we propose a patch-based convolutional neural network approach with a relatively small number of trainable parameters for COVID-19 diagnosis. 
The proposed method is inspired by our statistical analysis of the potential imaging biomarkers of the CXR radiographs. Experimental results show that our method achieves  state-of-the-art performance and provides clinically interpretable saliency maps, which are useful for COVID-19 diagnosis and patient triage.
\end{abstract}


\section{Introduction}
\label{sec:introduction}
\IEEEPARstart{C}{oronavirus} disease 2019 (COVID-19), caused by severe acute respiratory syndrome
coronavirus 2 (SARS-CoV-2),  has become global pandemic in less than four months since it was first reported,  reaching a 3.3 million confirmed cases and
238,000 death as of May 2nd, 2020.
Due to its highly contagious nature and lack of appropriate treatment and vaccines,
early detection of COVID-19 becomes increasingly important to prevent further spreading and to flatten
the curve for proper allocation of limited medical resources.

Currently, reverse transcription polymerase chain reaction  (RT-PCR), which
detects viral nucleic acid, is the golden standard for COVID-19 diagnosis, but
 RT-PCR results  using nasopharyngeal  and throat swabs can be affected by sampling errors and low viral load \cite{xie2020chest}.  Antigen tests may be fast, but have poor sensitivity. 
 
 Since most COVID-19 infected
patients were diagnosed with pneumonia, radiological
examinations may be useful for  diagnosis and assessment of disease progress.
Chest computed tomography (CT) screening on initial patient presentation showed outperforming sensitivity to RT-PCR \cite{fang2020sensitivity}
and even confirmed COVID-19 infection on negative or weakly-positive RT-PCR cases \cite{xie2020chest}. 
Accordingly,  recent COVID-19 radiological literature primarily focused on CT findings \cite{ai2020correlation,fang2020sensitivity}.
However, as the prevalence of COVID-19
increases,  the routine use of CT  places a huge burden on radiology
departments and potential infection of the CT suites; so the need to recognize COVID-19 features
on chest X-ray (CXR) is increasing. 

Common chest X-ray findings reflect those described by CT such as bilateral,
peripheral consolidation and/or ground glass opacities \cite{ai2020correlation,fang2020sensitivity}.
Specifically,
Wong et al \cite {wong2020frequency} described frequent chest X-ray (CXR) appearances on COVID-19.  
Unfortunately, it is reported that chest X-ray findings have a lower sensitivity than initial RT-PCR testing (69\%
versus 91\%, respectively) \cite{wong2020frequency}.
Despite this low sensitivity,  CXR abnormalities were
detectable in 9\% of patients whose initial RT-PCR was negative.

As the COVID-19 pandemic
threatens to overwhelm healthcare systems worldwide, 
CXR may be considered as a tool for
identifying COVID-19 if the diagnostic performance with CXR is improved. 
Even if CXR cannot completely replace the RT-PCR,  the indication of pneumonia is a clinical
manifestation of patient at higher risk requiring hospitalization, so
CXR
can be used for patient triage, determining the priority of patients' treatments  to help saturated healthcare system in the pandemic situation. 
 {This is especially important, since the most frequent known etiology of community acquired pneumonia is bacterial infection in general  \cite{brown1998community}. By excluding these population by triage, limited medical resource can be spared substantially.}

\begin{figure*}[!t]
\centerline{\includegraphics[width=2\columnwidth]{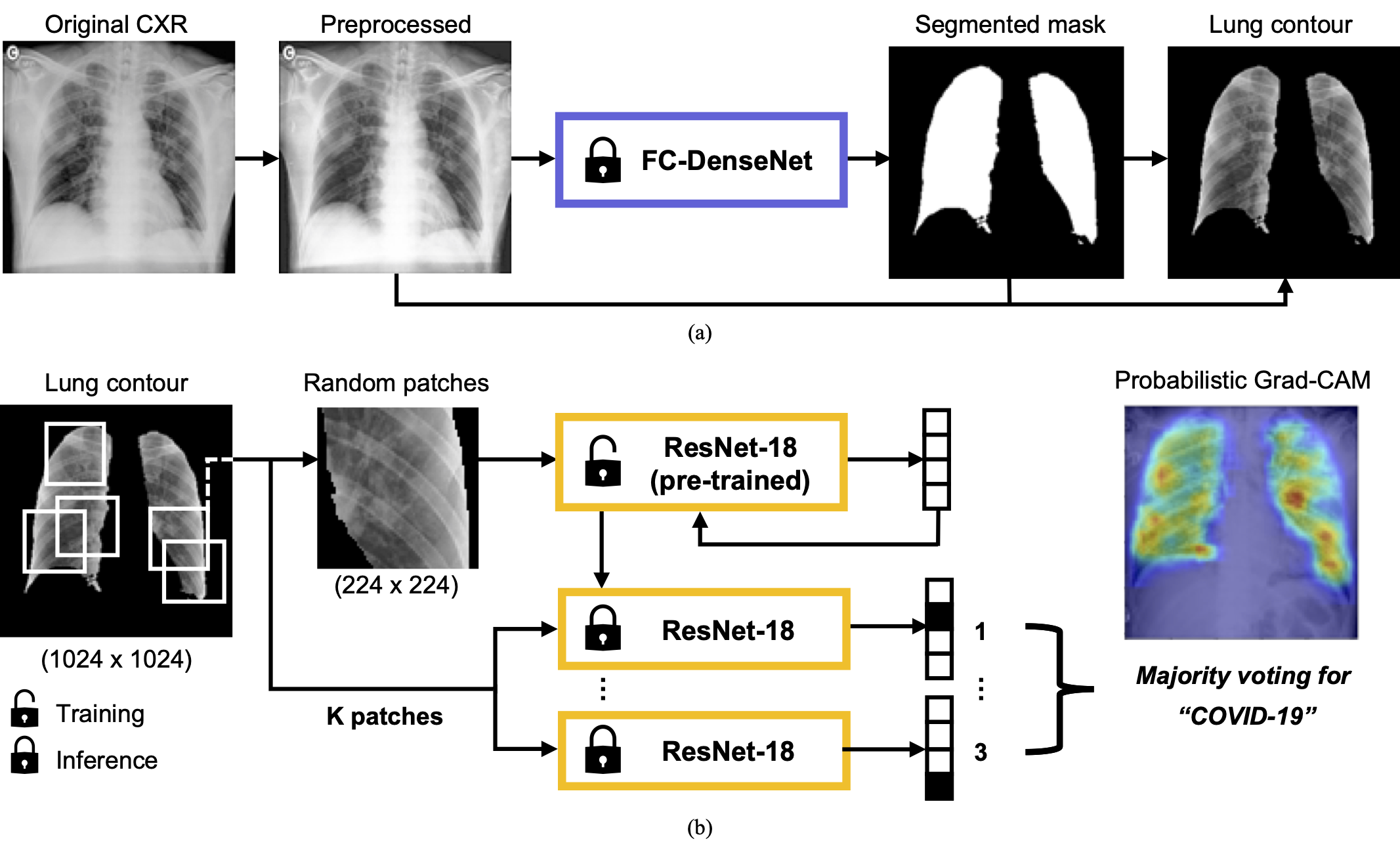}}
\caption{Overall architecture of the proposed neural network approach: (a) Segmentation network, and (b) Classification network  }
\label{fig_network}
\end{figure*}

Accordingly, deep learning (DL) approaches on chest X-ray for COVID-19 classification have been actively explored \cite{wang2020covid,narin2020automatic,apostolopoulos2020covid,hemdan2020covidx,apostolopoulos2020extracting,farooq2020covid,afshar2020covid}.
Especially,  Wang et al \cite{wang2020covid} proposed
 an open source deep convolutional neural network platform called COVID-Net that is 
 tailored for the detection of COVID-19 cases from
chest radiography images. 
They claimed that COVID-Net can
achieve good sensitivity for COVID-19 cases with 80\% sensitivity.

Inspired by this early success, in this paper
we aim to further investigate deep convolutional
neural network and evaluate its feasibility for COVID-19 diagnosis.
Unfortunately, under the current public health emergency, it is difficult to collect large set of
well-curated data for training neural networks. Therefore, one of the main focuses of this paper is to develop
a neural network architecture that is suitable for training  with limited training data set, which can still produce radiologically interpretable results.
Since most frequently observed distribution patterns of COVID-19 in  CXR are bilateral involvement, peripheral distribution and ground-glass opacification (GGO) \cite {salehi2020coronavirus}, a properly
designed neural network should reflect such radiological findings.

To achieve this goal, we first investigate several  imaging biomarkers that are often used in CXR analysis, such as lung area intensity distribution, 
the cardio-thoracic ratio, etc.  Our analysis found that there are statistically significant differences in the patch-wise intensity distribution, which is well-correlated with the radiological findings of the localized intensity variations in COVID-19 CXR.
This findings lead us to propose a novel patch-based deep neural network architecture with random patch cropping, from which the final
classification result are obtained by majority voting from inference results at multiple patch locations.
One of the important advantages of the proposed method is that due to the patch training the network complexity is relative small
and  {multiple patches in each image can be used to augment training data set,} so that even with the limited data set the neural network can be trained efficiently without overfitting.  {By combining with our novel preprocessing step to normalize the data heterogeneities and bias,}
we demonstrate that the proposed network architecture provides better sensitivity and interpretability, compared to the existing COVID-Net \cite{wang2020covid}  with the same data set.
 
 {
Furthermore, by extending the  idea of the gradient-weighted class activation map (Grad-CAM) \cite{selvaraju2017grad},
yet another important contribution of this paper is 
a novel 
probabilistic Grad-CAM that takes into account of patch-wise disease probability in generating global saliency map.}
The resulting class activation map clearly show the interpretable results that are well correlated with radiological findings.

\section{Proposed Network Architecture}


The overall algorithmic framework is given in  Fig. \ref{fig_network}.
The CXR images are first pre-processed for data normalization,
after which the pre-processed data are fed into   a segmentation network, from which
 lung areas can be extracted as shown in Fig. \ref{fig_network}(a).  
 From the segmented lung area, classification network is used to classify the corresponding diseases using
 a patch-by-patch training and inference, after which the final decision is made based on the
 majority voting  as shown in Fig.~\ref{fig_network}(b). 
  {Additionally, a probabilistic Grad-CAM saliency map is calculated to provide an interpretable result.}
 In the following, each network is described in detail.

\subsection{Segmentation network}

Our segmentation network aims to extract lung and heart contour from the chest radiography images. 
We adopted an extended fully convolutional (FC)-DenseNet103 to perform semantic segmentation \cite{jegou2017one}. The training objective is
\begin{equation}
\operatorname*{argmin}_\Theta \mathcal{L}(\Theta)
\label{eq_loss}
\end{equation}
where $\mathcal{L}(\Theta)$ is the cross entropy loss of multi-categorical semantic segmentation and $\Theta$ denotes the network parameter set, which is composed
of filter kernel weights and biases. Specifically, 
$\mathcal{L}(\Theta)$ is defined as
\begin{equation}
\mathcal{L}(\Theta) = -\sum_s\sum_j\lambda_{s} \mathds{1}(y_{j} = s){\log(p_{\Theta}(x_{j}))}
\label{eq_crossentropy}
\end{equation}
where $\mathds{1}(\cdot)$ is the indicator function,
$p_\Theta(x_{j})$ denotes the softmax probability of the $j$-th pixel in a CXR image $\xb$, and $y_{j}$ denotes the corresponding ground truth label. $s$ denotes class category, i.e., $s \in$ \{background, heart, left lung, right lung\}. $\lambda_{s}$ denotes weights given to each class category.

CXR images from different dataset resources may induce heterogeneity in their bits depth, compression type, image size, acquisition condition, scanning protocol,  postprocessing, etc.
Therefore, we develop a universal preprocessing step  {for data normalization} to ensure uniform intensity histogram throughout the entire dataset. The detailed preprocessing steps are as follows:
\begin{enumerate}
\item  {Data type casting (from uint8/uint16 to float32)} 
\item Histogram equalization (gray level = [0, 255.0])
\item Gamma correction ($\gamma$ = 0.5)
\item Image resize (height, width = [256, 256])
\end{enumerate}

Using the preprocessed data,  we trained \textit{FC-DenseNet103}\cite{jegou2017one} as our backbone segmentation network architecture. Network parameters were initialized by random distribution. We applied Adam optimizer \cite{kingma2014adam} with an initial learning rate of 0.0001. Whenever training loss did not improve by certain criterion, the learning rate was reduced by factor 10. We adopted early stopping strategy based on validation performance. Batch size was optimized to 2. We implemented the network using PyTorch library \cite{paszke2017automatic}.

\subsection{Classification network}
 {The classification network aims to classify the chest X-ray images according to the types of disease. We adopted the relatively
simple ResNet-18 as the backbone of our classification algorithm for two reasons. The first is to prevent from overfitting, since it is known that overfitting can occur when using an overly complex model for small number of data. Secondly, we intended to do transfer learning with pre-trained weights from ImageNet to compensate for the small training data set. 
We found that  these strategy make the training stable even when the dataset size is small.}

The labels were divided into four classes: normal, bacterial pneumonia, tuberculosis (TB), and viral pneumonia which includes the pneumonia caused by COVID-19 infection. We assigned the same class for viral pneumonia from other viruses (e.g. SARS-cov or MERS-cov) with COVID-19, since it is reported that  they have similar radiologic features even challenging for the experienced radiologists \cite{yoon2020chest}. Rather, we concentrated on more feasible work such as distinguishing bacterial pneumonia or tuberculosis from viral pneumonia, which show considerable differences in the radiologic features and are still useful for patient triage.

The pre-processed images were first masked with the lung masks from the segmentation networks, which are then fed into 
a classification network.
Classification network were implemented in  two different versions: global  and local approaches. 
In the global approach, the masked images were resized to $224 \times 224$, which were fed into the network.
 This approach is focusing on the global appearance of the CXR data, and was used as a baseline network for comparison.
 In fact, many of the existing researches employs similar procedure  \cite{wang2020covid,narin2020automatic,apostolopoulos2020covid,hemdan2020covidx}.

In the local patch-based approach, which is our proposed method, the masked images were cropped randomly with a size of $224 \times 224$, and resulting patches were used as the network inputs as shown in Fig. \ref{fig_network}(b). 
In contrast to the global approach,  {various CXR images are resized to a much bigger  $1024 \times 1024$ image  for our classification network 
to reflect the original pixel distribution better.} 
 Therefore,  the segmentation mask from Fig. \ref{fig_network}(a) are upsampled to match the  $1024 \times 1024$ image size.
To avoid cropping the patch from the empty area of the masked image, the centers of patches were randomly selected within the lung areas. During the inference, $K$-number of patches were randomly acquired  for each image  to represent the entire attribute of the whole image. The number $K$ was chosen to sufficiently cover all lung pixels multiple times.
Then, each patch was fed into the network to generate network output, and among $K$ network output
the final decision was made based on majority voting, i.e.  the most frequently declared class were regarded as final output  as depicted in Fig. \ref{fig_network}(b).
In this experiments, the number of random patches $K$ was set to 100, which means that 100 patches were generated randomly from one whole image for majority voting.

For network training,
  {pre-trained parameters from ImageNet are used for network weight initialization},
 after which the network was trained using the CXR data. As for optimization algorithm, Adam optimizer \cite{kingma2014adam} with learning rate of 0.00001 was applied. The network were trained for 100 epochs, but we adopted early stopping strategy based on validation performance metrics. The batch size of 16 was used. We applied weight decay and $L_1$ regularization to prevent overfitting problem. 
The classification network was also implemented by Pytorch library.

\subsection{Probabilistic Grad-CAM saliency map visualization}
 {We investigate the interpretability of our approach by visualizing a  saliency map.
One of the most widely used saliency map visualization methods is so-called gradient weighted class activation map (Grad-CAM) \cite{selvaraju2017grad}.
Specifically, the Grad-CAM saliency map of
the class $c$ for a given input image $\xb\in \Rd^{m\times n}$ is defined by }
\begin{align}\label{eq:cam}
\lb^{c}(\xb)=\textsc{Up}\left(\sigma\left(\sum_{k}\alpha_k^c \fb^k(\xb) \right)\right)\in \Rd^{m\times n}
\end{align}
 {where $\fb^k(\xb)\in \Rd^{u\times v}$ is the $k$-th feature channel at the last convolution layer (which corresponds to the layer 4 of ResNet-18 in our case), $\textsc{Up}(\cdot)$ denotes the upsampling operator from a $u\times v$ feature map to the $m\times n$ image, $\sigma(\cdot)$ is 
the rectified linear unit (ReLU) \cite{selvaraju2017grad}. Here,
 $\alpha_k^c$ is the feature weighted parameter for the class $c$, which can be obtained by}
\begin{align}
\alpha_k^c=\frac{1}{Z}\sum_{i=1}^{uv} \frac{\partial y_c}{\partial f_{i}^k}
\end{align}
 {for some scaling parameter $Z$, where $y_c$ is the score for the class $c$ before the softmax layer and $f_{i}^k$ denotes the $i$-th pixel value of $\fb^k(\xb)$. The Grad-CAM map $\lb^{c}$ is then normalized to have value in $[0,1]$.
The Grad-CAM for the global approach is used  as a baseline saliency map.}

 {However, care should be taken in applying Grad-CAM  to our local patch-based approach, since each patch has different score for the COVID-19 class. 
Therefore,  to obtain the global saliency map, patch-wise Grad-CAM saliency maps should be weighted with the estimated probability of the class, and their average value should be computed.
More specifically, our probabilistic Grad-CAM with respect to the input image $\xb \in \Rd^{m\times n}$
has the following value at the $i$-th pixel location:}
\begin{align}\label{eq:pcam}
\left[\lb_{prob}^{c}(\xb)\right]_{i}=  \frac{1}{K_{i}}\left[\sum_{k=1}^{K} r^c(\xb_k) \Qc_k\left( \lb^{c}(\xb_k)\right)\right]_{i}
\end{align}
 {where 
$\xb_k\in \Rd^{p\times q}$ denotes the  $k$-th input patch, $\Qc_k:\Rd^{p\times q}\mapsto \Rd^{m\times n}$ refers to the
operator that copies the $p\times q$-size $k$-th patch into the appropriate location of a zero-padded image in $\Rd^{m\times n}$,
 and  $\lb^{c}(\xb_k)\in \Rd^{p\times q}$ denotes the Grad-CAM  computed by \eqref{eq:cam}
with respect to the input patch $\xb_k\in \Rd^{p\times q}$,  and  $K_{i}$ denotes the number of the  frequency of the $i$-th pixel in the total $K$ patches.
Additionally, the class probability  $r^c(\xb_k)$ for the $k$-th patch can be readily calculated after the soft-max layer.
Accordingly,   the average probability of each pixel belonging to a  given class
 can be taken into consideration in Eq.~\eqref{eq:pcam} when constructing a global saliency map.}

\section{Method}

\subsection{Dataset}

We used public CXR datasets, whose characteristics are summarized in Table \ref{tab_dataset} and Table \ref{tab_dataset_class}.
 In particular, the data  in Table \ref{tab_dataset}  are used for training and validation of the segmentation networks, since the
 ground-truth segmentation masks are available.
 The curated data in  Table \ref{tab_dataset_class}  are from some of the data in   Table \ref{tab_dataset}  as well as other COVID-19 resources, which were used for training, validation, and test for the classification network.
 More detailed descriptions of the dataset are follows.

\subsubsection{Segmentation network dataset}

The JSRT dataset was released by the Japanese Society of Radiological Technology (JSRT) \cite{Shiraishi2000JSRT}. Total 247 chest posteroanterior (PA) radiographs were collected from 14 institutions including normal and lung nodule cases. Corresponding segmentation masks were collected from the SCR database \cite{van2006segmentation}. The JSRT/SCR dataset were randomly split into training (80\%) and validation (20\%). 
For cross-database validation purpose, we used another public CXR dataset: U.S. National Library of Medicine (USNLM) collected Montgomery Country (MC) dataset \cite{jaeger2014two}. Total 138 chest PA radiographs were collected including normal, tuberculosis cases and corresponding lung segmentation masks.

\begin{table} [!h]
\centering
\caption{Segmentation dataset resources}
\setlength{\tabcolsep}{3pt}
\begin{tabular}{ l l c c c c }
\hline
\multirow{2}{*}{\textbf{Dataset}}& \multirow{2}{*}{\textbf{Class}}& \multirow{2}{*}{\textbf{\#}}& \multirow{2}{*}{\textbf{Bits}}& \multicolumn{2}{c}{\textbf{Mask}}\\
& & & & \textbf{Lung}& \textbf{Heart}\\
\hline
\hline
\multicolumn{2}{l}{\textbf{Training}}&&&&\\
JSRT/SCR\cite{Shiraishi2000JSRT}\cite{van2006segmentation}& Normal/Nodule& 197& 12 & O & O\\
\hline
\multicolumn{2}{l}{\textbf{Validation}}&&&&\\
JSRT/SCR& Normal/Nodule& 50& 12 & O & O\\
NLM(MC)\cite{jaeger2014two}& Normal& 73& 8 & O & -\\
\hline
\end{tabular}
\label{tab_dataset}
\end{table}

\begin{table} [!h]
\centering
\caption{Classification data set resources}
\setlength{\tabcolsep}{3pt}
\begin{tabular}{ l l c c c c }
\hline
\multirow{2}{*}{\textbf{Dataset}}& \multirow{2}{*}{\textbf{Class}}& \multirow{2}{*}{\textbf{\#}}& \multirow{2}{*}{\textbf{Bits}}& \multicolumn{2}{c}{\textbf{Mask}}\\
& & & & \textbf{Lung}& \textbf{Heart}\\
\hline
\hline
JSRT/SCR& Normal& 20& 12 & O & O\\
NLM(MC)& Normal& 73& 8 & O & -\\
CoronaHack\cite{coronahack2020cor}& Normal& 98& 24 & - & -\\
NLM(MC)& Tuberculosis& 57& 8 & O & -\\
CoronaHack& Pneumonia (Bacteria)& 21& 24 & - & -\\
\multirow{3}{*}{Cohen et al\cite{cohen2020covid}}& Pneumonia (Bacteria)& 33& 24 & - & -\\
& Pneumonia (Virus)& 20& 24 & - & -\\
& Pneumonia (COVID-19)& 180& 24 & - & -\\
\hline
\end{tabular}
\label{tab_dataset_class}
\end{table}

\subsubsection{Classification dataset}

The dataset resources for the classification network are described in Table~\ref{tab_dataset_class}.
Specifically,
for  normal cases, the JSRT dataset and the NLM dataset from the segmentation validation dataset were included. 
For comparing COVID-19 from normal and different lung diseases, data were also collected from different sources \cite{cohen2020covid, coronahack2020cor}, including additional normal cases. 
 These datasets were selected because they are fully accessible to any research group, and they provide the labels with detailed diagnosis of disease. This enables more specific classification of pneumonia into bacterial and viral pneumonia, which should be classified separately because of their distinct clinical and radiologic differences. 

In the collected  data from the public dataset\cite{coronahack2020cor}, over 80\% was pediatric CXR from Guangzhou Women and Children’s Medical Center \cite{kermany2018identifying}. Therefore, to avoid the network from learning biased features from age-related characteristics, we excluded pediatric CXR images.  {This is because we aim to 
utilize CXR radiography with 
unbiased age distribution for more accurate evaluation of deep neural networks for COVID-19 classification. }


\begin{table} [th]
\centering
\caption{Disease class  summary of the data set}
\setlength{\tabcolsep}{3pt}
\begin{tabular}{c c c c c c c}
\hline
\textbf{Dataset}& \textbf{Normal} & \textbf{Bacterial} & \textbf{Tuberculosis} & \textbf{Viral} & \textbf{COVID-19} & \textbf{Total}\\
\hline
\hline
\textbf{Training} & 134 & 39 & 41 & 14 & 126 & 354\\
\textbf{Validation} & 19 & 5 & 5 & 2 & 18 & 49\\
\textbf{Test} & 38 & 10 & 11 & 4 & 36 & 99\\
\hline
\textbf{Total} & 191 & 54 & 57 & 20 & 180 & 502\\
\hline
\end{tabular}
\label{tab_classification_dataset}
\end{table}

Total dataset was curated into five classes; normal, tuberculosis, bacterial pneumoia, viral pneumonia, COVID-19 pneumonia. 
 The numbers of each disease class from the data set are summarized in Table \ref{tab_classification_dataset}. Specifically, a total of 180 radiography images of 118 subjects from COVID-19 image data collection were included. 
 Moreover, 
 a total of 322 chest radiography images from different subjects were used, which include 191, 54, and 20 images for normal, bacterial pneumonia, and viral pneumonia (not including COVID-19), respectively. The combined dataset were randomly split into train, validation, and test sets with the ratio of 0.7, 0.1, and 0.2.
 

\subsubsection{Dataset for comparison with COVID-Net}
We prepared a separate dataset to compare our method with existing state-of-the art (SOTA) algorithm called COVID-Net \cite{wang2020covid}. COVID-19 image data collection was combined with RSNA Pneumonia Detection Challenge dataset as described in \cite{wang2020covid}  for a fair comparison between our method and COVID-Net. 
The reason we separately train our network with the COVID-Net data set is that 
RSNA Pneumonia Detection Challenge dataset provide only the information regarding the presence of pneumonia, rather than the detailed diagnosis of disease, so that  the labels were divided into only three categories including normal, pneumonia, and COVID-19 as in Table \ref{tab_COVIDNet_dataset}. More specifically, there were 8,851 normal and 6,012 pneumonia chest radiography images from 13,645 patients in RSNA Pneumonia Detection Challenge dataset, and these images were combined with COVID-19 image data collection to compose a total dataset. Among these, 100 normal, 100 pneumonia, and 10 COVID-19 images were randomly selected for validation and test set, respecitvely as in \cite{wang2020covid}.
Although we believe our categorization into  normal, bacterial, TB, and viral+COVID-19 cases is more correlated with the radiological findings and practically useful in clinical environment \cite{yoon2020chest}, we conducted this  additional comparison experiments with the data set in Table \ref{tab_COVIDNet_dataset} to demonstrate that our algorithm provides competitive performance compared to COVID-Net in the same experiment set-up.

\begin{table} [th]
\centering
\caption{Dataset for comparison with COVID-Net}
\setlength{\tabcolsep}{3pt}
\begin{tabular}{c c c c c}
\hline
\textbf{Dataset}& \textbf{Normal} & \textbf{Pneumonia} & \textbf{COVID-19} & \textbf{Total}\\
\hline
\hline
\textbf{Training} & 8651 & 5812 & 160 &  {14623}\\
\textbf{Validation} & 100 & 100 & 10 & 210\\
\textbf{Test} & 100 & 100 & 10 & 210\\
\hline
\textbf{Total} & 8851 & 6012 & 180 & 15043\\
\hline
\end{tabular}
\label{tab_COVIDNet_dataset}
\end{table}

\subsection{Statistical Analysis of Potential CXR COVID-19 markers}

The following standard biomarkers from CXR image analysis are investigated.

\begin{figure}[!hbt]
\centerline{\includegraphics[width=\columnwidth]{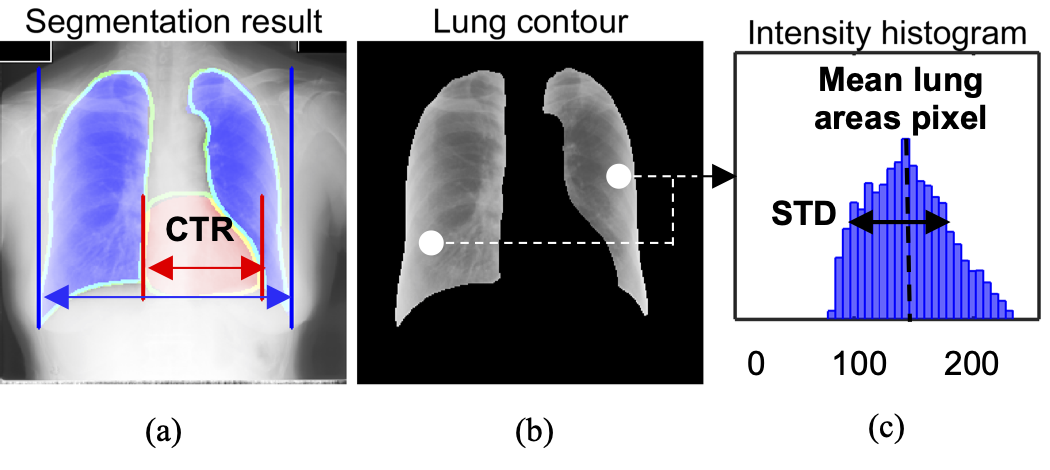}}
\caption{(a) Segmentation result. Each lung and heart segment are overlapped on CXR coloring in blue and red, respectively. Green line represent the ground truth. (b) Extracted lung areas, and (c) corresponding lung area pixel intensity histogram. }
\label{fig_marker_intro}
\end{figure}

\begin{itemize}
\item {\em Lung morphology:}
Morphological structures of the segmented lung area as illustrated in Fig. \ref{fig_marker_intro}(b) was evaluated throughout different classes.
\item {\em Mean lung intensity:}
From the segmented lung area, we calculated mean value of the pixel intensity within the lung area as shown in Fig.~\ref{fig_marker_intro}(c). 
\item {\em Standard deviation of lung intensity:}
From the intensity histogram of lung area pixels, we calculated one standard deviation which is indicated as the black double-headed arrow in Fig.~\ref{fig_marker_intro}(c). 
\item {\em Cardiothoracic Ratio (CTR):}
CTR can be calculated by dividing the maximal transverse cardiac diameter by the maximal internal thoracic diameter annotated repectively as red and blue double-headed arrows in Fig.~\ref{fig_marker_intro}(a). 
Cardiothoracic Ratio (CTR) is  a widely used marker to diagnosis cardiomegaly \cite{danzer1919cardiothoracic, dimopoulos2013cardiothoracic}. We hypothesized that if cardiothoracic boundary become blurred by rounded opacities or consolidation in COVID-19 CXR \cite{ai2020correlation,fang2020sensitivity,wong2020frequency}, distinct off-average CTR value can be utilized as an abnormality alarm. 
\end{itemize}

Statistical analysis for the potential biomarkers was performed using MATLAB 2015a (Mathworks, Natick). Kolmogorov Smirnov test was used to evaluate the normal distribution of marker candidates. For non-normally distributed variables, Wilcoxon signed rank test was used to compare segmentation performance with identical data size, and Wilcoxon rank sum test was used to compare COVID-19 marker candidates to those of other classes with different data sizes. Statistical significance (SS) levels were indicated as asterisks; * for $p<$0.05, ** for $p<$0.01 and *** for $p<$0.001.

\subsection{Classification performance metrics}
The performance of the classification methods was evaluated using the confusion matrix. From the confusion matrix, true positive (TP), true negative (TN), false positive (FP), and false negative (FN) values were obtained, and 5 metrics for performance evaluation were calculated as below:
\begin{enumerate}
\item $\mathrm{Accuracy} = (TN + TP) / (TN + TP + FN + FP) $
\item $\mathrm{Precision} = TP / (TP + FP) $
\item $\mathrm{Recall} = TP / (TP + FN) $
\item $\mathrm{F1\,score} = 2 (\mathrm{Precision} \times \mathrm{Recall})/ (\mathrm{Precision} + \mathrm{Recall}) $
\item $\mathrm{Specificity} = TN / (TN + FP) $ 
\end{enumerate}
Among these, the F1 score was used as the evaluation metric for early stopping. The overall metric scores of the algorithm were calculated by averaging each metric for multiple classes.

%
%

\section{EXPERIMENTAL RESULTS}
\label{experimental results}

\subsection{Segmentation performance on cross-database}
Segmentation performance of anatomical structure was evaluated using Jaccard similarity coefficient.  Table \ref{tab_segmentation_result} presents the Jaccard similarity coefficient of each contour on the validation dataset. The results confirmed  {our method provides
comparable accuracy to previous works using 
the JSRT dataset and the NLM(MC) dataset \cite{candemir2019review, jangam2018public}}.

\begin{table}[!h]
\centering
\caption{CXR segmentation results}
    \begin{tabular}{lcccc}
    \toprule
    \multirow{2}[2]{*}{} & \multirow{2}{*}{\textbf{Preprocess}} & \multicolumn{3}{c}{\textbf{Jaccard similarity coefficient}} \\
         &      & \textbf{Lung} & \textbf{SS} & \textbf{Heart} \\
    \midrule
    \midrule
    \textbf{JSRT} & O    & 0.955$\pm$0.015 &      & 0.889$\pm$0.054 \\
    \textbf{NLM(MC)} & -    & 0.932$\pm$0.022 & \multirow{2}[1]{*}{***} &  \\
    \textbf{NLM(MC)} & O    & 0.943$\pm$0.013 &      &  \\
    \bottomrule
    \end{tabular}%
\label{tab_segmentation_result}%
\end{table}%

To evaluate segmentation performance on cross-database, we tested either original or preprocessed images of the NLM dataset as  inputs. The result shows that our universal preprocessing step for data normalization  contributes to the processing of cross-database with statistically significant improvement on segmentation accuracy (Jaccard similarity coefficients from 0.932 to 0.943,  $p<$ 0.001). This result indicates that preprocessing is crucial factor to ensure segmentation performance in cross-database.

\subsection{Morphological analysis of lung area}
\begin{figure}[!ptb]
\centerline{\includegraphics[width=1\columnwidth]{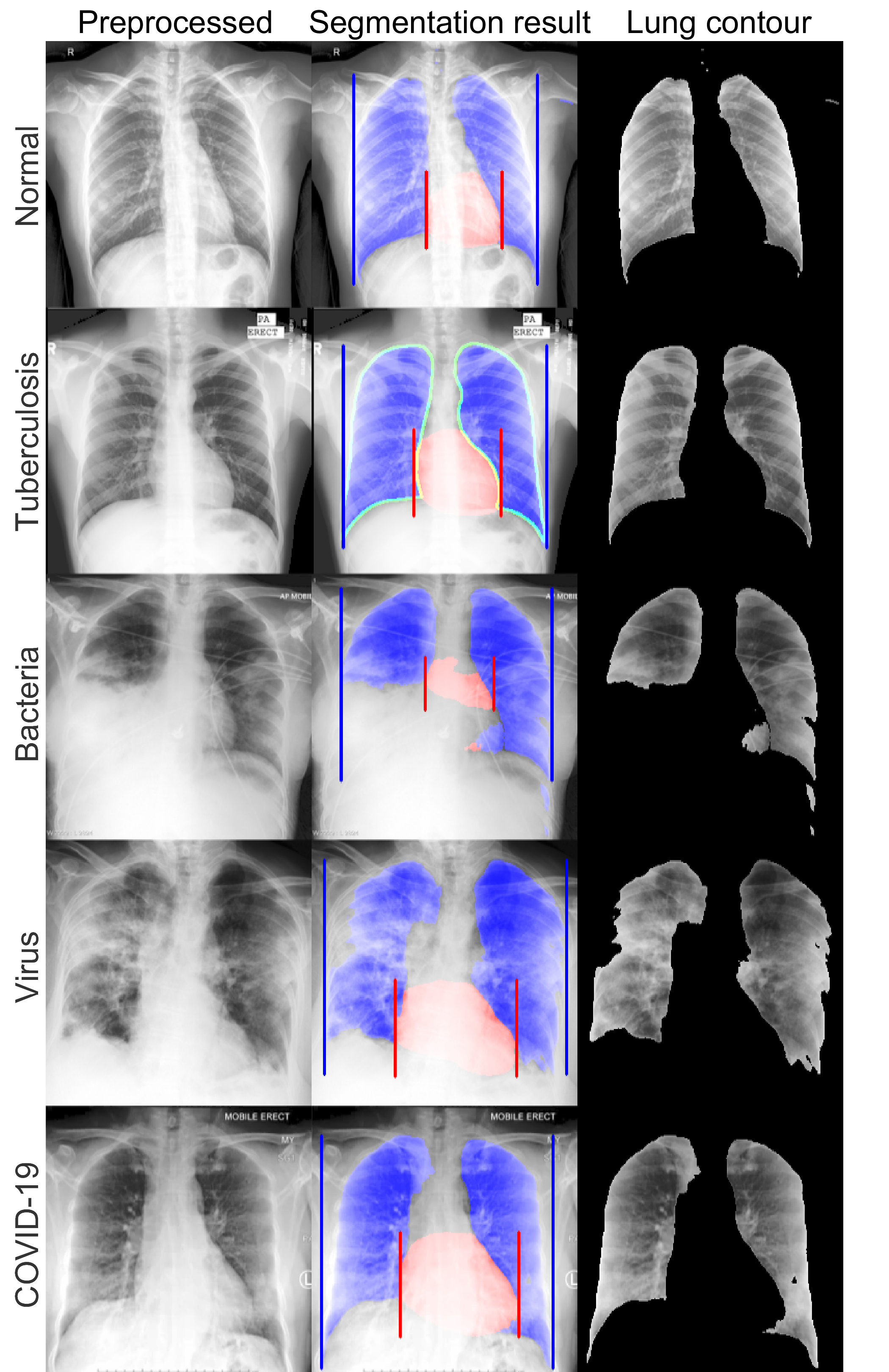}}
\caption{Preprocessed images, corresponding segmentation results, and the extracted lung contours are shown along with the column-axis. Each row depicts different categorical class.}
\label{fig_classes_result}
\end{figure}

To analyze morphological characteristics in the segmented lung area, a representative CXR radiograph for each class was selected for
visual evaluation. Lung contour of each class showed differentiable features and showed mild tendency. In normal and tuberculosis cases (the first and the second row of Fig. \ref{fig_classes_result}, respectively), overall lung and heart contour were well-segmented. 
In the bacterial case, however, the segmented lung area was deformed due to wide spread opacity of bacterial pneumonia as shown in the third row of Fig. \ref{fig_classes_result},  and both the right cardiac and thoracic borders were lost. 
In overall bacterial infection cases, similar findings were occasionally observed which caused degraded segmentation performance.
 {This suggests that abnormal morphology of the segmentation masks may be a useful biomarker to \added{differentiate the severe}
   infections.}
In the fourth row of Fig. \ref{fig_classes_result}, viral infection caused bilateral consolidations \cite{paul2004radiologic}, thus partial deformation of lung area was observed. In the COVID-19 case of the fifth row of  Fig. \ref{fig_classes_result},
 despite the bi-basal infiltrations \cite{thevarajan2020breadth}, lung area was fully segmented. In overall cases of the viral and the COVID-19 classes, lung areas were either normally or partially-imcompletely segmented,
 {so morphological features of the segmentation masks may not be  sufficiently discriminatory markers for viral and COVID-19 classes.}
 Based on these morphological findings in segmented lung area, we further  {investigated other} potential COVID-19 biomarkers.

\begin{figure}[!htb]
\begin{subfigure}{.5\textwidth}
\centering
{\includegraphics[width=0.8\columnwidth]{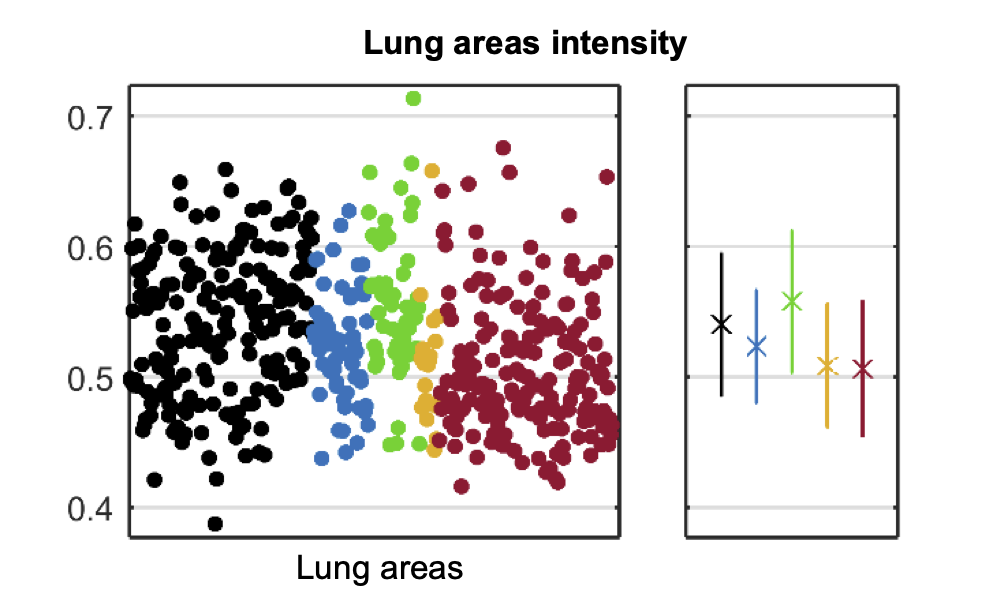}}
\caption{}
\label{fig_param1}
\end{subfigure}
\begin{subfigure}{.5\textwidth}
\centering
{\includegraphics[width=0.8\columnwidth]{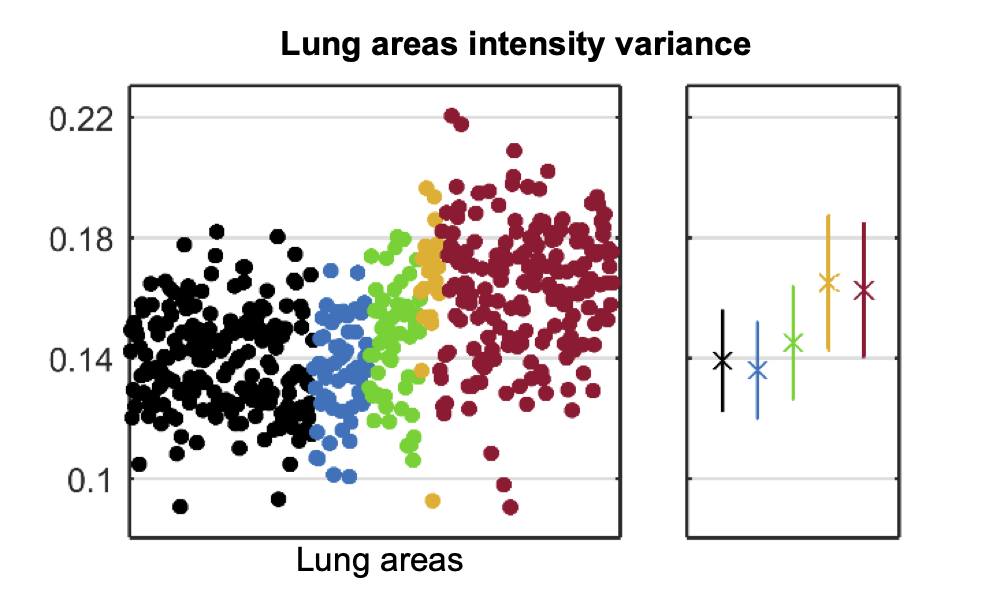}}
\caption{}
\label{fig_param2}
\end{subfigure}
\begin{subfigure}{.5\textwidth}
\centering
{\includegraphics[width=0.8\columnwidth]{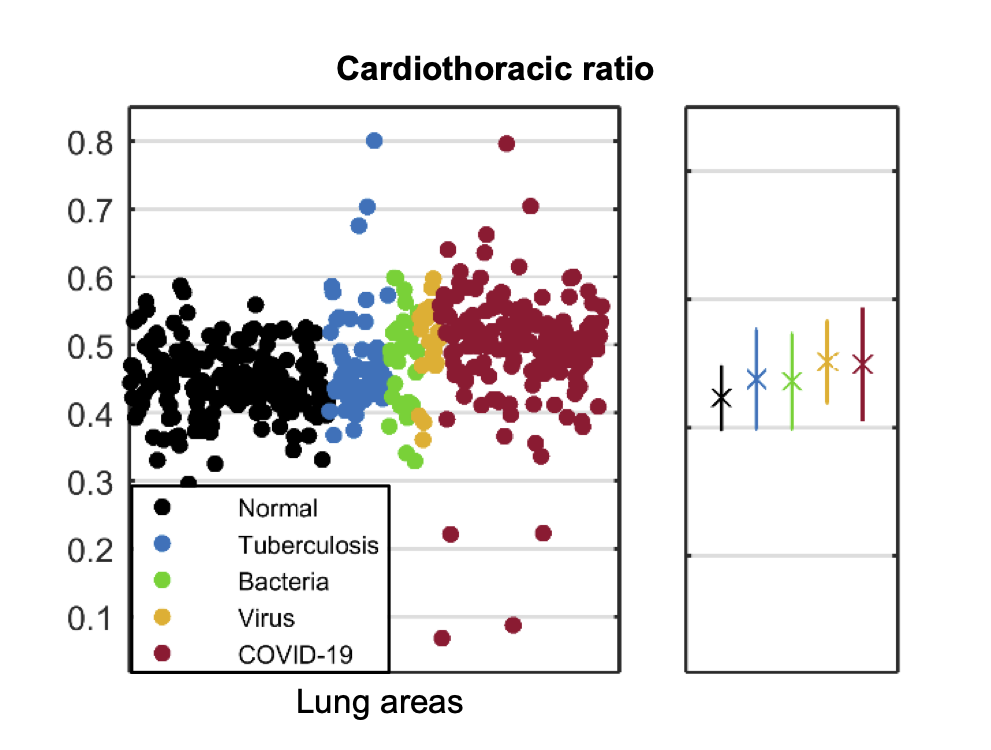}}
\caption{}
\label{fig_param3}
\end{subfigure}
\caption{Scatter plot (Left) and corresponding mean values with one standard deviation error bars (Right). All the parameter values are normalised to an arbitrary unit.}
\label{fig_scatter}
\end{figure}

\subsection{Statistical significancy of potential COVID-19 bio-markers}
\label{marker_feasibility}

 {We  hypothesized that 
CXR appearance influenced by consolidations or infiltration of COVID-19  may be reflected in intensity of the radiograph.
Thus, 
intensity-related COVID-19 marker candidates were extracted and compared. }

\subsubsection{\paramone{}}
Mean pixel intensity of each lung area is shown in the scatter plot of Fig. \ref{fig_scatter}(a). COVID-19 cases showed lower mean intensity compared to other cases with statistical significance level ($p<$0.001 for normal and bacterial, $p<$0.01 for TB). 
Table \ref{tab_param1} describes the corresponding statistical result.
Despite the statistical significance, the scatter plot showed broad overlap between several classes. 

\begin{table}[!h]
  \centering
  \caption{\paramone\, statistics}
    \begin{tabular}{lllllll}
    \toprule
         & \multicolumn{1}{c}{\multirow{2}{*}{\textbf{Mean}}} & \multicolumn{1}{c}{\multirow{2}{*}{\textbf{STD}}} & \multicolumn{4}{c}{\textbf{Statistical significance}} \\
         &      &      & \textbf{Normal} & \textbf{TB} & \textbf{Bacteria} & \textbf{Viral} \\
    \midrule
    \midrule
    \textbf{Normal} & 0.540  & 0.055  &      &      &      &  \\
    \textbf{Tuberculosis} & 0.523  & 0.043  & *    &      &      &  \\
    \textbf{Bacteria} & 0.558  & 0.054  & -    & ***  &      &  \\
    \textbf{Viral} & 0.509  & 0.047  & **   & -    & ***  &  \\
    \textbf{COVID-19} & 0.506  & 0.051  & ***  & **   & ***  & - \\
    \bottomrule
    \end{tabular}%
  \label{tab_param1}%
\end{table}%

\subsubsection{\paramtwo{}}
Standard deviation of pixel intensity of each lung area are scattered in plot in Fig. \ref{fig_scatter}(b). For both the COVID-19 and the viral cases, the variance values were higher than other classes with statistical significance ($p<$0.001 for all). 
Table \ref{tab_param2} describes the corresponding statistical result. 

\begin{table}[!h]
  \centering
  \caption{\paramtwo\, statistics}
     \begin{tabular}{lllllll}
    \toprule
         & \multicolumn{1}{c}{\multirow{2}{*}{\textbf{Mean}}} & \multicolumn{1}{c}{\multirow{2}{*}{\textbf{STD}}} & \multicolumn{4}{c}{\textbf{Statistical significance}} \\
         &      &      & \textbf{Normal} & \textbf{TB} & \textbf{Bacteria} & \textbf{Viral} \\
    \midrule
    \midrule
    \textbf{Normal} & 0.139  & 0.017  &      &      &      &  \\
    \textbf{Tuberculosis} & 0.136  & 0.016  & -    &      &      &  \\
    \textbf{Bacteria} & 0.145  & 0.019  & *    & **   &      &  \\
    \textbf{Viral} & 0.165  & 0.022  & ***  & ***  & ***  &  \\
    \textbf{COVID-19} & 0.163  & 0.022  & ***  & ***  & ***  & - \\
    \bottomrule
    \end{tabular}%
  \label{tab_param2}%
\end{table}%

\begin{table}[!h]
\centering
\caption{\paramtwo{} statistics by excluding AP supine radiographs}
    \begin{tabular}{lllllll}
    \toprule
         & \multicolumn{1}{c}{\multirow{2}{*}{\textbf{Mean}}} & \multicolumn{1}{c}{\multirow{2}{*}{\textbf{STD}}} & \multicolumn{4}{c}{\textbf{Statistical significance}} \\
         &      &      & \textbf{Normal} & \textbf{TB} & \textbf{Bacteria} & \textbf{Viral} \\
    \midrule
    \midrule
    \textbf{Normal} & 0.139  & 0.017  &      &      &      &  \\
    \textbf{Tuberculosis} & 0.136  & 0.016  & -    &      &      &  \\
    \textbf{Bacteria} & 0.143  & 0.020  & -    & -    &      &  \\
    \textbf{Viral} & 0.163  & 0.025  & ***  & ***  & **   &  \\
    \textbf{COVID-19} & 0.161  & 0.022  & ***  & ***  & ***  & - \\
    \bottomrule
    \end{tabular}%
\label{tab_ablation}%
\end{table}%

To investigate the effect of scanning protocol on statistics, we performed  additional study by excluding AP Supine radiographs from entire dataset with documented patient information. Recall that AP Supine protocol is an alternative to standard PA or AP protocol depending on patient condition. Since AP Supine protocol is not common in normal cases, supine scanning with different acquisition condition may have potential for considerable heterogeneity in data distribution, causing biased results in statistical analysis, so we investigated this issue.
The result shown in Table \ref{tab_ablation} compared to Table \ref{tab_param2} showed minor difference in both the COVID-19 and the viral and cases. 
The result indicates that for both the COVID-19 and viral classes, the highly intensity-variable characteristic in the lung area is invariant to scaning protocol.

\subsubsection{\paramthree{}}
CTR values of each lung area is scattered in Fig. \ref{fig_scatter}(c). Despite there exist statistical differences between the COVID-19 cases to other classes ($p<$0.001 for normal and TB, $p<$0.05 for Bacteria), the scatter plot showed broad overlap between several classes. Table \ref{tab_param3} describes the corresponding statistical result.

\begin{table}[!h]
  \centering
  \caption{\paramthree{} statistics}
    \begin{tabular}{lllllll}
    \toprule
         & \multicolumn{1}{c}{\multirow{2}{*}{\textbf{Mean}}} & \multicolumn{1}{c}{\multirow{2}{*}{\textbf{STD}}} & \multicolumn{4}{c}{\textbf{Statistical significance}} \\
         &      &      & \textbf{Normal} & \textbf{TB} & \textbf{Bacteria} & \textbf{Viral} \\
    \midrule
    \midrule
    \textbf{Normal} & 0.446  & 0.051  &      &      &      &  \\
    \textbf{Tuberculosis} & 0.476  & 0.078  & *    &      &      &  \\
    \textbf{Bacteria} & 0.472  & 0.074  & -    & -    &      &  \\
    \textbf{Viral} & 0.502  & 0.064  & ***  & *    & -    &  \\
    \textbf{COVID-19} & 0.499  & 0.086  & ***  & ***  & *    & - \\
    \bottomrule
    \end{tabular}%
  \label{tab_param3}%
\end{table}%

Based on the statistical analysis of  potential bio-marker candidates, we found that
 intensity distribution pattern within the lung area may be most effective in the diagnosis,
 which highly reflects the reported chest X-ray (CXR) appearances of COVID-19, i.e., multi-focally distributed consolidation and GGO in specific region such as peripheral and lower zone \cite{ai2020correlation,fang2020sensitivity,wong2020frequency}. 

However, care should be taken, since not only the
locally concentrated multiple opacities can cause uneven intensity distribution throughout entire lung area, but also
different texture distribution within CXR may cause the similar intensity variations.
For example, multi-focally distributed consolidation from 
COVID-19 could make the intensity variance differentiating factor from other classes, but also  bacterial pneumonia generates opacity as well, whose feature may lead to the similar intensity distributions  as results of different characteristics of opacity spreading pattern.

\begin{figure}[!hbt]
\begin{subfigure}{.5\textwidth}
\centering
{\includegraphics[width=0.8\columnwidth]{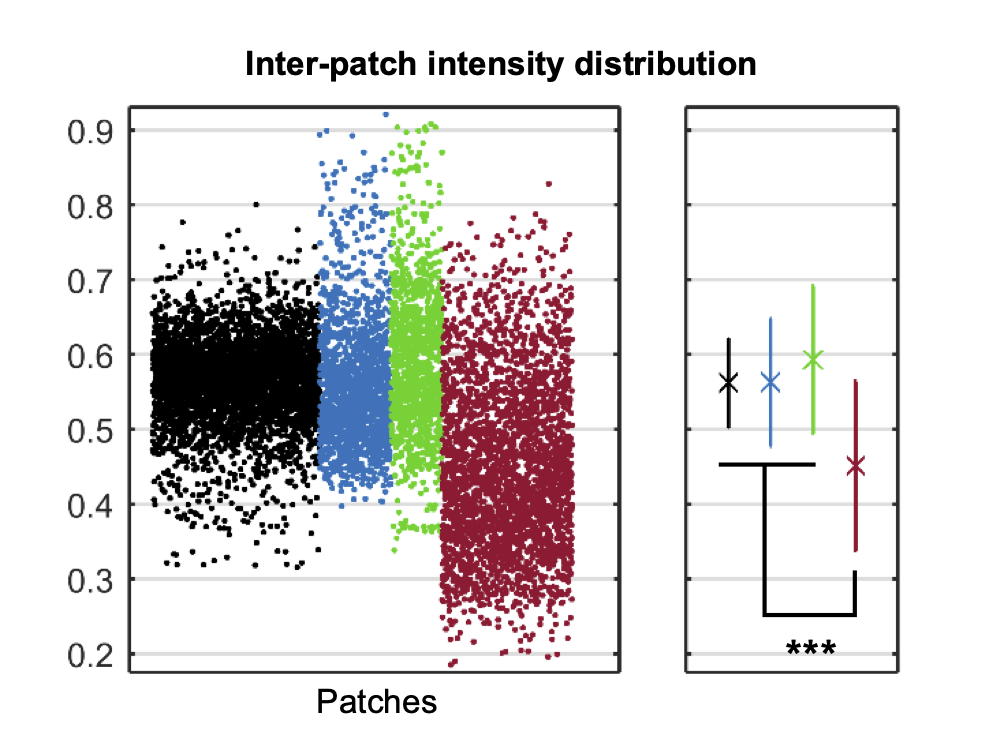}}
\caption{}
\label{fig_patch_inter}
\end{subfigure}
\begin{subfigure}{.5\textwidth}
\centering
\includegraphics[width=0.8\columnwidth]{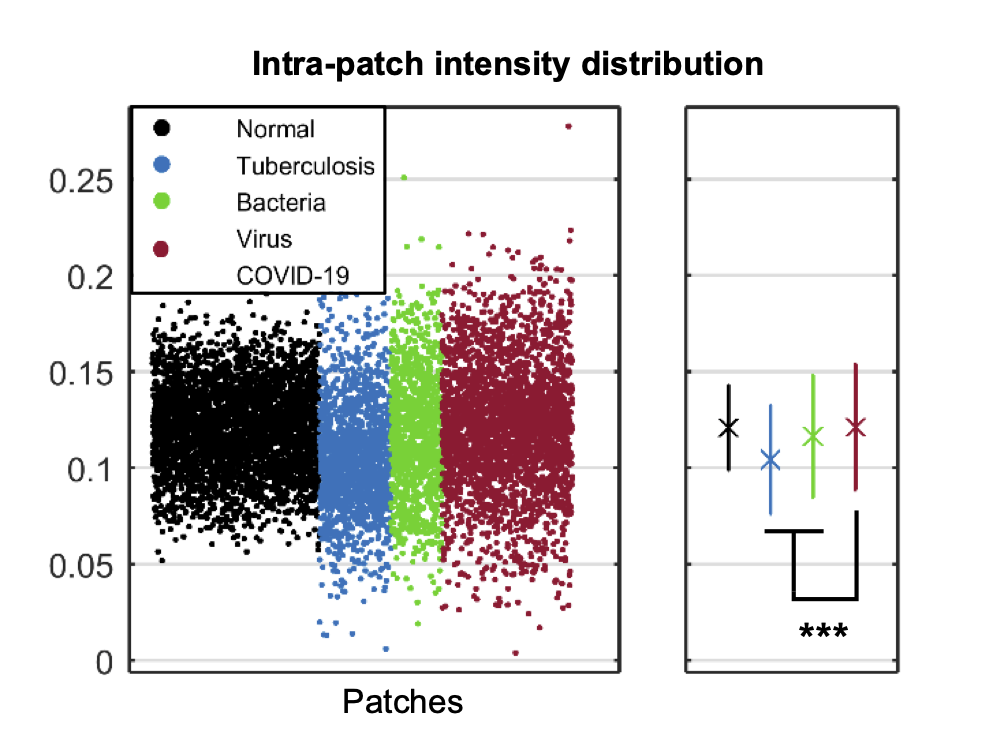}
\caption{}
\label{fig_patch_intra}
\end{subfigure}
\caption{Scatter plot (Left) and corresponding mean values with one standard deviation error bars. Each scatter depicts a patch which was correctly classified to the ground truth label. All the parameter values were normalised to an arbitrary unit. 
Statistically differentiable classes from the COVID-19 and viral cases ($p<$0.001) are marked at each error bar. }
\label{fig_patch_scatter}
\end{figure}

To decouple these compounding effects,
 we further investigated the local and global
   intensity distribution. For the correctly classified patches from our classification network, we computed their mean intensity and standard deviation (STD) values.
  We refer to the distribution of mean intensity of each patch as the inter-patch intensity distribution (Fig. \ref{fig_patch_scatter}(a)) and the STD of each patch as intra-patch intensity distribution (Fig. \ref{fig_patch_scatter}(b)). 
As shown in Fig. \ref{fig_patch_scatter}(a),
the inter-patch intensity distribution of the unified COVID-19 and viral class showed distict lower intensity values ($p<$0.001 for all) to other classes and highly intensity-variant characteristics which can be represented as the large error bar. This result is in accordance with the result of lung area intensity and intensity variance (Fig. \ref{fig_scatter}(a), (b)). Intra-patch intensity distribution, however, showed no difference compared to the normal class ($p>$0.05). From these intra- and inter-patch intensity distribution results, we can infer that 
intra-patch variance, which represents local texture information, was not crucially informative, whereas
the globally distributed multi-focal intensity change may be an important discriminating feature for COVID-19 diagnosis, which
is strongly correlated with the radiological findings.


One common finding among the marker candidates was no difference between the COVID-19 and the viral case ($p>$0.05 for all the markers), which is also correlated with radiological findings \cite{yoon2020chest}.
 Therefore, in the classification network, the COVID-19 and viral classes were integrated into one class.

\subsection{Classification performance}

\begin{figure}[!hbt]
\centerline{\includegraphics[width=0.73\columnwidth]{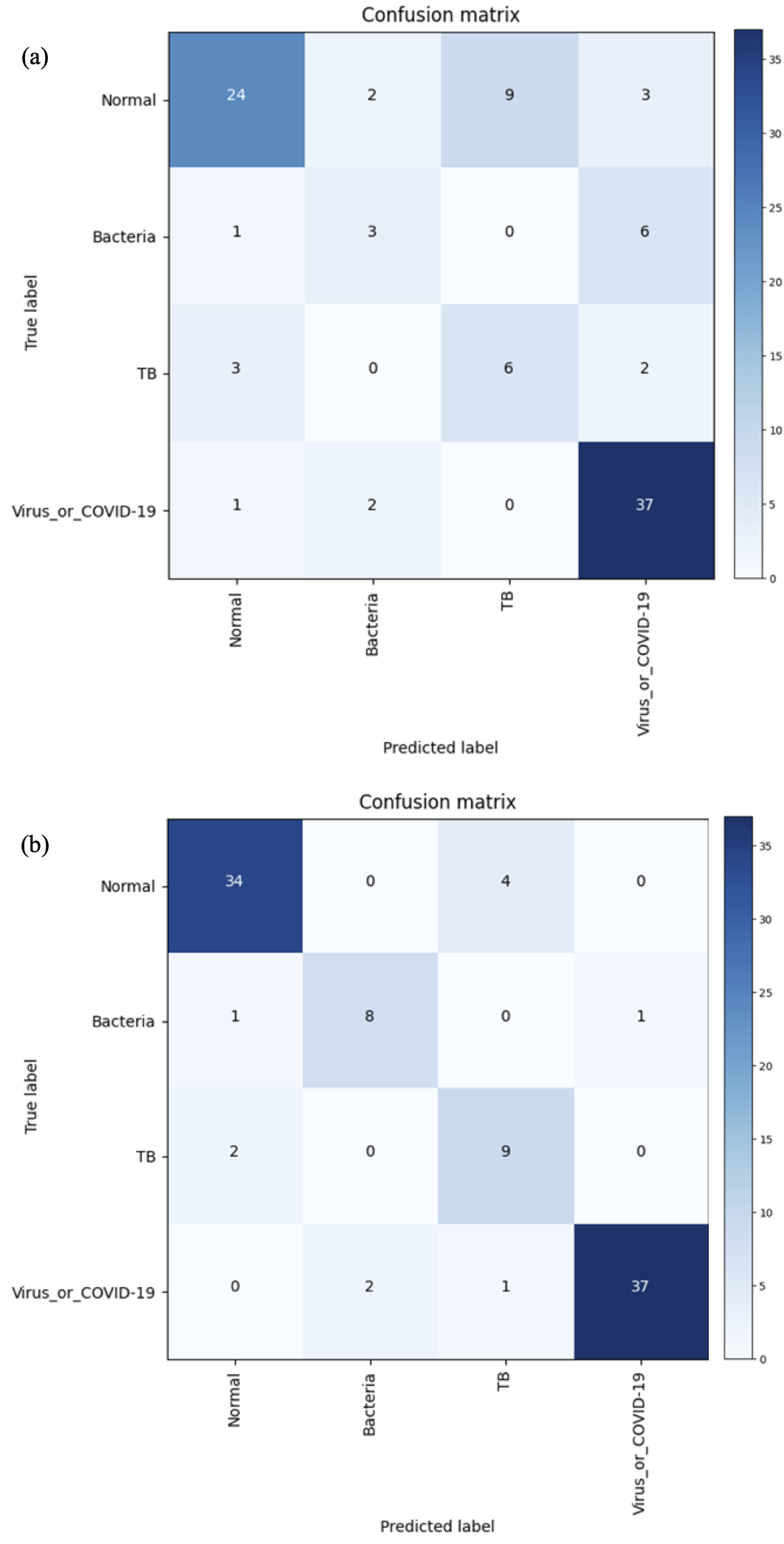}}
\caption{Confusion matrices for the (a) global approach, and (b) the proposed local patch-based approach.}
\label{figure_cm}
\end{figure}

The classification performances of the proposed method are provided in Table \ref{tab_classify_performance}. The confusion matrices for the (a) global method and the (b) local patch-based method are shown in Fig. \ref{figure_cm}. The proposed local patch-based approach showed consistently better performance than global approach in all metrics. 
  {In particular, as depicted in Table~\ref{tab_comparison_sensitivity},
our method showed the sensitivity of $92.5\%$ for COVID-19 and viruses, which was acceptable performance as a screening method, considering the fact that the sensitivity of COVID-19 diagnosis by X-ray image is known to be $69\%$ even for clinical experts and that the current gold standard, RT-PCR, has sensitivity of $91\%$ \cite{wong2020frequency}. Moreover, compared to the global approach, the sensitivity of other classes
are significantly high, which confirms the efficacy of our method.}

\begin{table} [!hbt]
\centering
\caption{Classification results from the global approach and the proposed patch-based classification network.}
\setlength{\tabcolsep}{3pt}
\resizebox{\columnwidth}{!}{\begin{tabular}{ c c c c c c}
\hline
\textbf{Methods}& \textbf{Accuracy} & \textbf{Precision} & \textbf{Recall} & \textbf{F1 score} & \textbf{Specificity}\\
\hline
\hline
\textbf{Global approach} & 70.7 & 60.6 & 60.1 & 59.3 & 89.7\\
\textbf{Local approach} & 88.9 & 83.4 & 85.9 & 84.4 & 96.4 \\
\hline
\end{tabular}}
\label{tab_classify_performance}
\end{table}

\begin{table} [th]
\centering
\caption{Sensitivity of the global approach and the local patch-based classification network}
\setlength{\tabcolsep}{3pt}
\resizebox{\columnwidth}{!}{\begin{tabular}{ c | c c c c }
\hline
\multirow{2}{*}{\textbf{Methods}}& & & \textbf{Sensitivity} \\ & \textbf{Normal} & \textbf{bacterial} & \textbf{Tuberculosis} & \textbf{COVID-19 and Viral} \\
\hline
\hline
\textbf{Global approach} & 63.2 & 30 & 54.5 & 92.5 \\
\textbf{Local approach} & 89.5 & 80 & 81.8 & 92.5 \\
\hline
\end{tabular}}
\label{tab_comparison_sensitivity}
\end{table}

\begin{figure}[!hbt]
\centerline{\includegraphics[width=1\columnwidth]{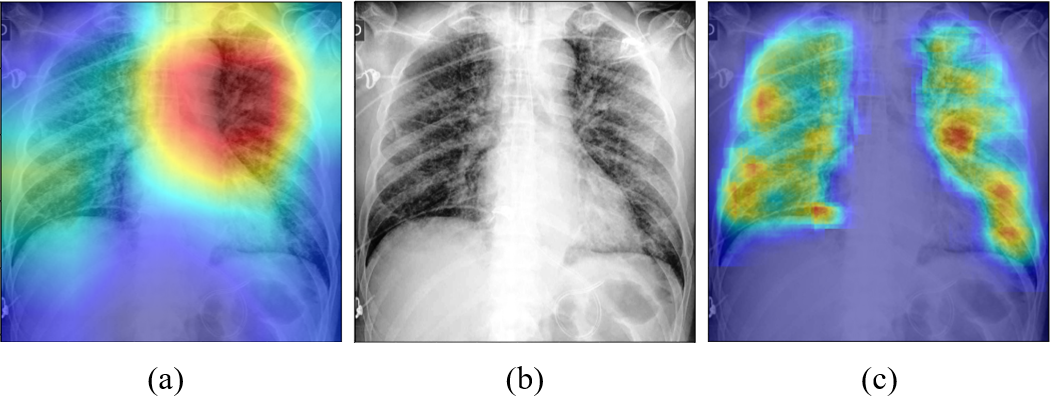}}
\caption{ {Examples of saliency maps for COVID-19 patient. (a) Grad-CAM  saliency map for the global approach, (b) original X-ray image, and (c) our  probabilistic Grad-CAM saliency map for local patch-based approach.}}
\label{fig_saliency_1}
\end{figure}

\begin{figure*}[!hbt]
\centerline{\includegraphics[width=15cm]{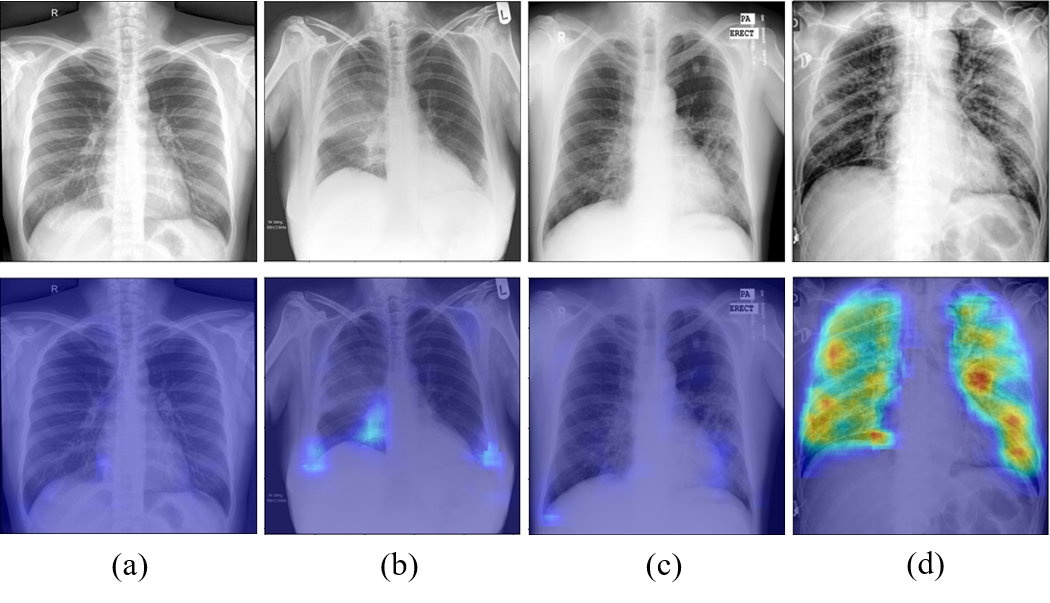}}
\caption{ {Examples of probabilistic Grad-CAM of COVID-19 class  for  (a) normal, (b) bacterial, (c) tuberculosis, and (d) COVID-19 pneumonia patients.}}
\label{fig_saliency_2}
\end{figure*}

\subsection{Interpretability  using   {saliency map}}

 {Fig. \ref{fig_saliency_1} and Fig. \ref{fig_saliency_2} illustrate the examples of visualization of  saliency map. As shown in Fig. \ref{fig_saliency_1}(a), the existing Grad-CAM method for global approach showed the limitation that it only focuses on the broad main lesion so that it cannot properly differentiate multifocal lesions within the image. On the other hand, with the probabilistic Grad-CAM, multifocal GGOs and consolidations were visualized effectively by our local patch-based approach as shown in Fig. \ref{fig_saliency_1}(c), which was in consistent with the findings reported by clinical experts. 
In particular, when we compute the probabilistic Grad-CAM for the COVID-19 class using patient images from various classes (i.e., normal, bacterial, TB, and COVID-19), a noticeable activation map was observed only in the COVID-19 patient data set, whereas almost no activations were observed in patients with other diseases and conditions as shown in Fig. \ref{fig_saliency_2}. These results strongly support our claim that the  probabilistic Grad-CAM saliency map  from our local patch-based approach is more intuitive and interpretable compared to the existing methods.}



\section{Discussion}

%
%
\subsection{COVID-19 Features on CXR}


%

In the diagnosis of COVID-19,
other diseases mimicking COVID-19 pneumonia should be differentiated, including
community-acquired pneumonia such as streptococcus pneumonia, mycoplasma and
chlamydia related pneumonia, and other coronavirus infections.

 In radiological literature, most frequently observed distribution patterns of COVID-19 are bilateral involvement, peripheral distribution and ground-glass opacification (GGO) \cite {salehi2020coronavirus}. 
Wong et al \cite {wong2020frequency} found that consolidation was the most common finding (30/64, 47\%),
followed by GGO (21/64, 33\%). CXR abnormalities had a peripheral (26/64, 41\%) and lower
zone distribution (32/64, 50\%) with bilateral involvement (32/64, 50\%), whereas pleural effusion was
uncommon (2/64, 3\%). 

Our statistical analysis of the intensity distribution clearly showed that the
globally distributed localized intensity variation is a discriminatory factor for COVID-19 CXR images,
which was also confirmed with our
saliency map. This 
clearly confirmed that the proposed method clearly reflects the radiological findings.

\subsection{Feasibility as a `triage' for COVID-19}
In pandemic situation of infectious disease, the distribution of medical resources is a matter of the greatest importance. As COVID-19 is spreading rapidly and surpassing the capacity of medical system in many countries, it is necessary to make reasonable decision to distribute the limited resources based on the `triage', which determine the needs and urgency for each patient. 
 {In general, the most common  cause of community acquired pneumonia is bacterial infection \cite{brown1998community}. Specifically, most studies reported that S. pneumoniae is the most frequent causative strain ($15-42\%$), after which H. influenza ($11-12\%$) and viral pneumonia follow as the second and the third most common causes of pneumonia, respectively  \cite{brown1998community}. In addition, depending on the geological region, substantial proportion of pneumonia may be diagnosed as tuberculosis (up to $10\%$) \cite{brown1998community}. Summing these up, the proportion of bacterial pneumonia and tuberculosis is suspected to be still large even in this pandemic situation of COVID-19.} In this respect, the disease such as bacterial pneumonia or tuberculosis as well as normal condition can be excluded primarily, to preserve limited medical resources such as RT-PCR or CT only for those who suspected to be infected with COVID-19.

\begin{figure}[!hbt]
\centerline{\includegraphics[width=1\columnwidth]{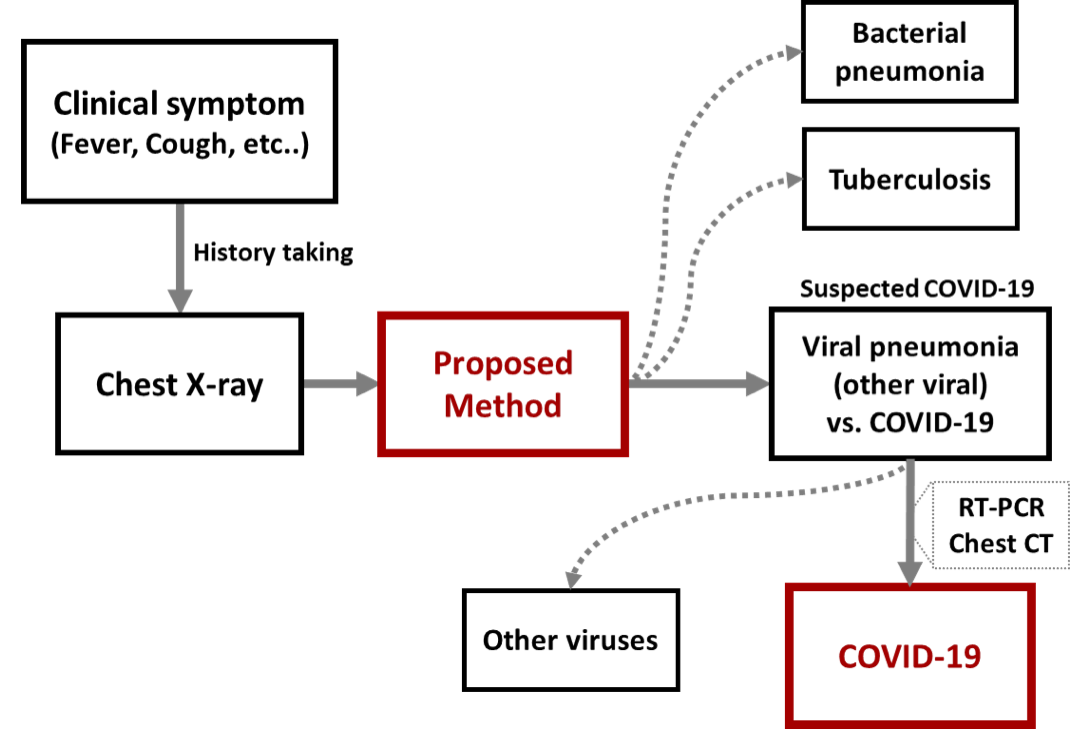}}
\caption{Potential triage workflow that utilizes the proposed algorithm in the diagnosis of COVID-19 patient.}
\label{fig_workflow}
\end{figure}

 The detailed triage workflow that utilizes the proposed algorithm is described in Fig. \ref{fig_workflow}.
Specifically, our neural network is trained to classify other viral and COVID-19 in the same class. This is not only because it is strongly
correlated with the radiological findings \cite{yoon2020chest}, but also useful as a triage. More specifically, by excluding normal,  bacterial pneumonia, and TB at the early stage, we can use RT-PCR  or CT for only those patients classified as other virus and COVID-19 cases for final diagnosis.
By doing this procedure,  we can save limited medical resources such as RT-PCR or CT to those patients whose diagnosis by CXR is difficult even by radiologists.

\subsection{Training stability}

In order to investigate  the origin of the apparent advantages of using local patch-based training over the global approach, we investigate
the training dynamics  to investigate the presence of overfitting. This is especially important, given that the training data is limited due to the difficulty of systematic CXR data collection for  COVID-19 cases under  current public health emergency.  

\begin{figure}[!hbt]
\centerline{\includegraphics[width=0.8\columnwidth]{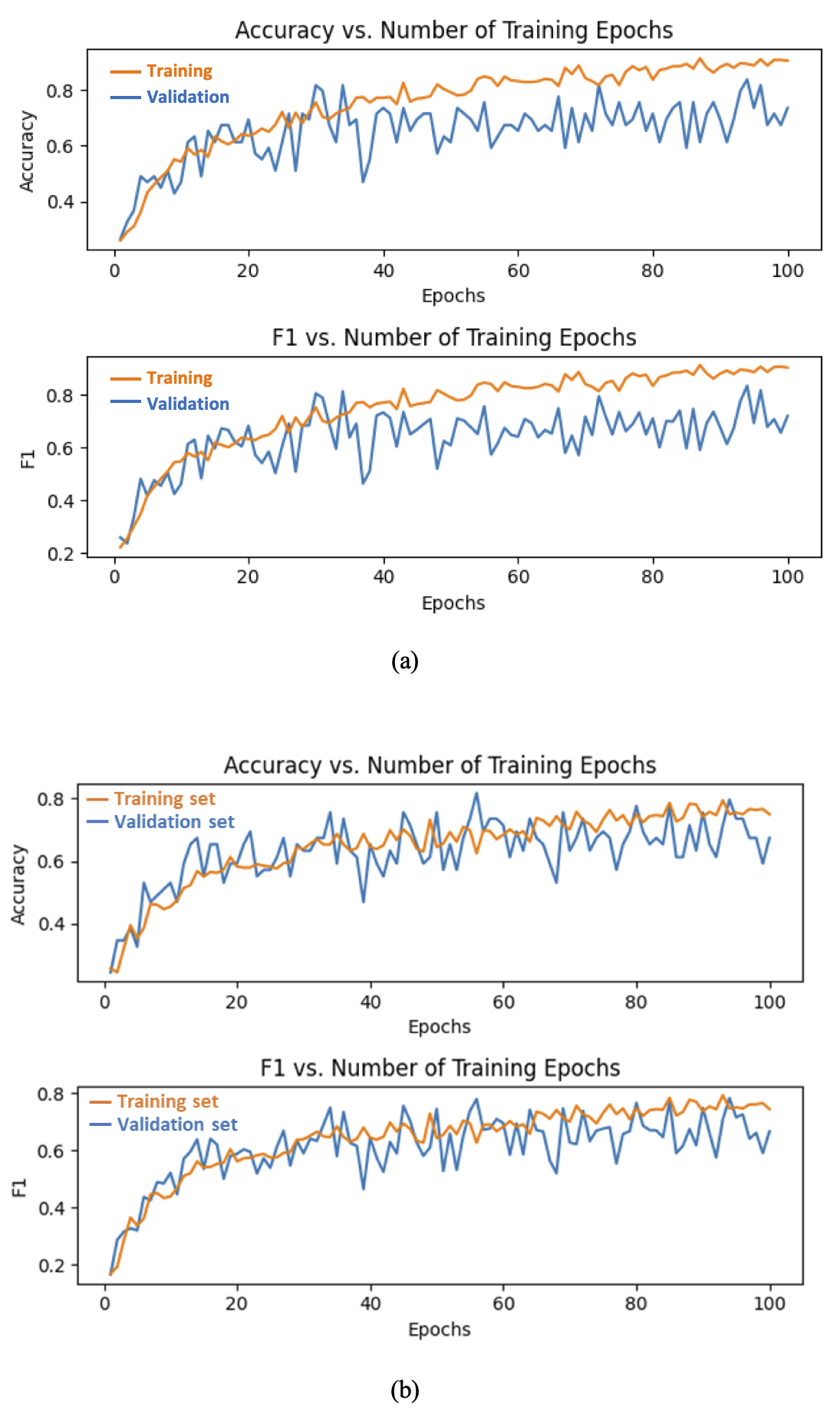}}
\caption{Training and validation accuracy and F1-score for each epoch. (a) Global approach, and (b) the proposed patch-based approach.}
\label{fig_curves}
\end{figure}

Fig. \ref{fig_curves} shows the curves for accuracy and F1 score of (a) the global approach and (b) the proposed local patch-based approach for each epoch. Note that both approaches use the same number of weight parameters.
Still, thanks to the increasing training data set from the random patch cropping across all image area, our local patch-based algorithm did not showed any sign of overfitting even with the small numbers of training data, while the global approach showed significant overfitting problem. This clearly indicates that with the limited data set the patch-based neural network training may be a promising direction.


\subsection{Comparison with COVID-Net}

Since the proposed patch-based neural network architecture is designed by considering limited training data set,
we investigated any potential performance loss in comparison with other state-of-the art  (SOTA) deep learning approach that has been developed without such consideration.
Specifically,  COVID-Net \cite{wang2020covid} is one of the most successful approaches in COVID-19 diagnosis, so we chose it as the SOTA method.

The comparison between our method and COVID-Net is shown in Table \ref{tab_comparison}. With the same dataset,
our method showed overall accuracy of 91.9 \% which is comparable to that of 92.4 \% for COVID-Net. 
Furthermore, our method provided significantly improved sensitivity to COVID-19 cases compared to the COVID-Net. 
In addition, it is also remarkable that our method uses only about 10\% number of parameters (11.6 M) compared to that of COVID-Net (116.6 M), because the proposed algorithm is developed based on less complex network architecture without increasing the complexity of the model. This may bring the advantages not only in the aspect of computational time but also in the aspect of performance and stability with small-sized dataset. 

\begin{table} [th]
\centering
\caption{Comparison of our method with COVID-Net}
\setlength{\tabcolsep}{3pt}
\resizebox{\columnwidth}{!}{\begin{tabular}{ c | c c c | c c c }
\hline
\multirow{2}{*}{\textbf{Methods}}& & \textbf{Sensitivity} & &&  \textbf{Precision} &\\ & \textbf{Normal} & \textbf{Pneumonia} & \textbf{COVID-19} & \textbf{Normal} & \textbf{Pneumonia} & \textbf{COVID-19}\\
\hline
\hline
\textbf{COVID-Net} & 95 & 91 & 80 & 91.3 & 93.8 & 88.9\\
\textbf{Proposed} & 90 & 93 & 100 & 95.7 & 90.3 & 76.9\\
\hline
\end{tabular}}
\label{tab_comparison}
\end{table}

\subsection{Cross-database generalization capability}

 {We are aware that
the current study has limitations due to the
lack of well-curated pneumonia CXR dataset.
Specifically,  our CXR data set come from a single or at most two sources (see Table \ref{tab_dataset_class}). 
 Moreover, publicly available COVID-19 dataset \cite{cohen2020covid} are largely extracted from online publications, website, etc, so they
 are not collected in a rigorous manner. } 
 
\begin{figure}[!tbh]
\centerline{\includegraphics[width=1\columnwidth]{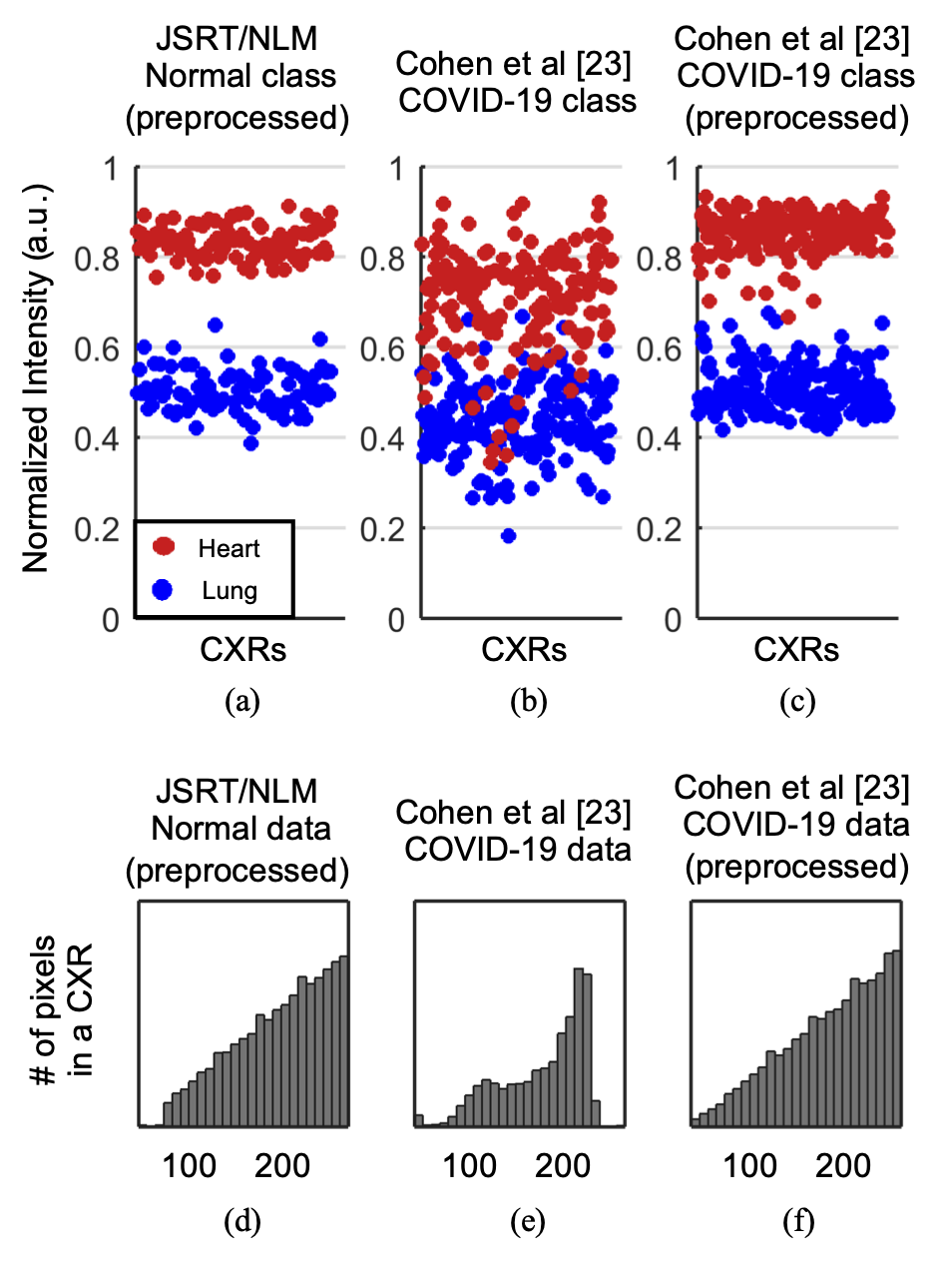}}
\caption{Intensity distribution of segmented anatomies of (a) normal, (b) original COVID-19, and (c) preprocessed COVID-19 CXRs. Representative intensity histogram of each (d) normal, (e) original COVID-19,  and (f) preprocessed COVID-19 CXR.}
\label{fig_covid_intensity}
\end{figure}

  {To mitigate the issue of potential bias from the limitation of the database, we employed
 a universal preprocessing step for data normalization  for the entire dataset as discussed before. 
We investigated the effects of our preprocessing step on cross-database generalization by investigating the COVID-19 dataset, which poses the most severe intra-dataset heterogeneity. As shown in Fig. \ref{fig_covid_intensity}(b), each original CXRs of the COVID-19 class showed highly-varying intensity characteristics among each segmented anatomies. Thanks to our preprocessing step, the mean pixel intensity distribution between lung and heart regions of  the preprocessed 
 COVID-19 dataset  (see Fig. \ref{fig_covid_intensity}(c)) became similar to the normal class in Fig. \ref{fig_covid_intensity}(a).
The problem of heterogeneity can be also mitigated  as shown in the intensity histograms (see Fig. \ref{fig_covid_intensity}(d)-(f)). 
The results confirmed that the original COVID-19 data could be well preprocessed to have comparable intensity distribution to that of well-curated normal data. 
}

\subsection{Segmentation network analysis}

\subsubsection{Comparison with U-Net}

 {Recall that we chose \textit{FC-DenseNet103} as a backbone segmentation network architecture thanks to its higher segmentation performance with smaller number of parameters (9.4 M)\cite{lobo2020applying}. To demonstrate the effectiveness of  CXR segmentation by the FC-Densenet103, we trained U-Net \cite{ronneberger2015u} under identical training conditions and compared the results. There was no significant difference between the networks result.

\begin{table}[H]
  \centering
  \caption{Lung segmentation results comparison}
    \begin{tabular}{l|lc|lc}
    \toprule
    \multicolumn{1}{l|}{\multirow{2}{*}{\textbf{Networks}}} & \multicolumn{4}{c}{\textbf{Lung jaccard similarity coefficient}} \\
         & \multicolumn{1}{c}{\textbf{JSRT}} & \textbf{SS} & \multicolumn{1}{c}{\textbf{NLM(MC)}} & \textbf{SS} \\
    \midrule
    \midrule
    \textbf{U-Net} & 0.955$\pm$0.014 & \multirow{2}{*}{$p>$0.05} & 0.941$\pm$0.014 & \multirow{2}{*}{\added{$p>$0.05}} \\
    \textbf{FC-DenseNet103} & 0.955$\pm$0.015 &      & 0.943$\pm$0.013 &  \\
    \bottomrule
    \end{tabular}%
  \label{tab_unet}%
\end{table}%

\added{We further analyzed the effect of the different segmentation methods on classification performance. The proposed segmentation method with the FC-DenseNet103 resulted consistently better classification performance in all metrics than the U-Net.} 
\added{When compared with the FC-DenseNet67, which has smaller number of parameters (3.5 M) \cite{jegou2017one},  the performance improvement by our method is significant.}
\added{
Given the better trade-off between the complexity versus performance, 
we adopted FC-DenseNet103 as our
segmentation network.}

\begin{table} [!hbt]
\centering
\caption{Classification results with different segmentation methods}
\setlength{\tabcolsep}{3pt}
\resizebox{\columnwidth}{!}{\begin{tabular}{ l| c c c c c}
\hline
\textbf{Netowrks}& \textbf{Accuracy} & \textbf{Precision} & \textbf{Recall} & \textbf{F1 score} & \textbf{Specificity}\\
\hline
\hline
\textbf{U-Net} & 85.9 & 82.3 & 84 & 82.5 & 95.3 \\
\textbf{FC-DenseNet67} & 81.8 & 73.1 & 76.6 & 74.3 & 91.5 \\
\textbf{FC-DenseNet103} & 88.9 & 83.4 & 85.9 & 84.4 & 96.4 \\
\hline
\end{tabular}}
\label{tab_classify_diff_segment}
\end{table}

\subsubsection{Effect of trainset size}

 {To demonstrate the robustness of the proposed segmentation network with limited training dataset, we performed the ablation study by reducing training dataset size. Lung segmentation performance was evaluated on the cross-database NLM(MC) dataset. For the preprocessed NLM(MC) dataset, Jaccard similarity coefficients remain stable until 50\% of trainset was used for training as shown in Table \ref{tab_seg_ablation}; however, in the original NLM(MC) dataset without preprocessing step, segmentation performance decreased steeply as the size of trainsets decreased. This results support that proposed segmentation network endures limited training dataset size by matching intensity distribution of cross-database CXRs thanks to our preprocessing step.}

\begin{table}[H]
  \centering
  \caption{Effects of  trainset size}
    \begin{tabular}{l|lc|lc}
    \toprule
    \multicolumn{1}{c|}{\multirow{3}{*}{\textbf{Trainset size}}} & \multicolumn{4}{c}{\textbf{Lung jaccard similarity coefficient}} \\
         & \multicolumn{1}{c}{\textbf{NLM }} & \textbf{SS } & \multicolumn{1}{c}{\textbf{NLM(MC) }} & \textbf{SS } \\
         & \multicolumn{1}{c}{\textbf{(preprocessed)}} & \textbf{100\%} & \multicolumn{1}{c}{\textbf{(original)}} & \textbf{100\%} \\
    \midrule
    \midrule
    \textbf{100\%} & 0.943$\pm$0.013 &      & 0.932$\pm$0.022 &  \\
    \textbf{50\%} & 0.941$\pm$0.018 & $p>$0.05 & 0.908$\pm$0.035 & *** \\
    \textbf{30\%} & 0.929$\pm$0.025 & ***  & 0.866$\pm$0.047 & *** \\
    \textbf{10\%} & 0.906$\pm$0.045 & ***  & 0.851$\pm$0.058 & *** \\
    \bottomrule
    \end{tabular}%
  \label{tab_seg_ablation}%
\end{table}%

\subsubsection{Segmentation effects on marker analysis and classification}

 {The proposed segmentation network was trained with normal subject set that have segmentation mask as shown  Table \ref{tab_segmentation_result}, \ref{tab_seg_ablation}, and showed comparable performance with the state-of-the-art for the normal subjects.
However, 
when  CXR images with severe consolidation are used, segmentation performance degradation is unavoidable, since such images have been never
observed during the training.

\added{For example, the radiograph from a bacterial pneumonia case  in Fig. \ref{fig_classes_result} was under-segmented
due to widely spread severe opacity.} 
\added{To further investigate this issue, we examined all cases of  under-segmentation  by defining the under-segmentation as a segmentation mask in which  over 1/4 of the entire lung region is deformed. Our investigation showed that the under-segmentations are only outliers that are observed  in some patient data set (7 of 54 bacterials cases (13\%,), 2 of 57 cases for TB (3.5\%), and 5 of 200 cases for COVID-19 and viral pneumonia (2.5\%)), 
whereas no under segmentation were observed from 191 healthy subjects.}

\added{To confirm that the difference in the segmentation results can be a morphological marker for classification, we evaluated whether it is possible to distinguish normal and abnormal images (including bacterial pneumonia, tuberculosis, COVID-19 and other viruses) using only binary segmentation masks (not X-ray images). With a separately trained neural network using only binary masks, it was possible to distinguish between normal and abnormal images with 86.9\% sensitivity. This confirms that the morphology of the segmentation mask is a discriminatory biomarker between normal and the patient groups.}
\added{Then, we  conducted additional experiments to evaluate how classification performance is affected by excluding or including under-segmentation cases.  Although the differences in other labels were not significant, the overall sensitivity for bacterial were better when the under-segmented subjects were excluded. 
Therefore, the under-segmentation still has some effects on the classification between the diseases.
}

 \added{
Finally, we performed an additional experiment for the comparison of classification with and without segmentation masks. The results in
Table \ref{tab_classify_without_mask} clearly confirmed that  despite the under-segmented outliers
the use of segmentation mask significantly improved the classification performance on the whole.
This suggests that there are  rooms to improve the performance
of the proposed method, if the segmentation network could be further trained using patient cases with correct segmentation labels.
}

%

\begin{table} [!hbt]
\centering
\caption{Classification results with and without segmentation mask}
\setlength{\tabcolsep}{3pt}
\resizebox{\columnwidth}{!}{\begin{tabular}{ c c c c c c}
\hline
\textbf{Methods}& \textbf{Accuracy} & \textbf{Precision} & \textbf{Recall} & \textbf{F1 score} & \textbf{Specificity}\\
\hline
\hline
\textbf{Without mask} & 79.8 & 72.7 & 59.7 & 60.7 & 91.9 \\
\textbf{With mask} & 88.9 & 83.4 & 85.9 & 84.4 & 96.4 \\
\hline
\end{tabular}}
\label{tab_classify_without_mask}
\end{table}

\subsection{Classification network analysis}
\subsubsection{Effect of patch size on the performance}

  {To evaluate the effect of the patch size on the performance of the classification algorithm, we tested various patch sizes, such as $112 \times 112$, and $448 \times 448$.  Using half-sized ($112 \times 112$) patches, the results were worse as shown in Table \ref{tab_patch}. With double-sized ($448 \times 448$) patches, the results were not better than those with patch size of $224 \times 224$, as depicted in Table \ref{tab_patch}. In summary, there seems to be a clear drawback in reducing the patch size, and there was also no benefit with increasing the patch size. Therefore, we chose the value in between.}

 {}
\begin{table} [!hbt]
\centering
\caption{Classification results using different patch sizes}
\setlength{\tabcolsep}{3pt}
\resizebox{\columnwidth}{!}{\begin{tabular}{ c c c c c c}
\hline
\textbf{Patch size}& \textbf{Accuracy} & \textbf{Precision} & \textbf{Recall} & \textbf{F1 score} & \textbf{Specificity}\\
\hline
\hline
\textbf{$112 \times 112$} & 78.8 & 71.2 & 75.9 & 72.3 & 93.2 \\
\textbf{$224 \times 224$} & 88.9 & 83.4 & 85.9 & 84.4 & 96.4 \\
\textbf{$448 \times 448$} & 84.8 & 78.8 & 80.1 & 79.3 & 94.8 \\
\hline
\end{tabular}}
\label{tab_patch}
\end{table}

\subsubsection{Effects of trainset size}

 {We analyzed the effect of dataset size in terms of classification performance, since we aimed to develop the method that has the advantage of maintaining robustness even when only limited data are available. The classification performances with decreasing dataset sizes are provided in Fig. \ref{fig_dataset_size}. The global approach using whole image, which is similar to most classification methods, showed prominent decrease in accuracy with decreasing dataset size, but the proposed local patch-based method showed less compromised performance, showing the robustness to the reduced  dataset size as shown in Fig. \ref{fig_dataset_size}(a). These results were more prominent when comparing them by relative scale as shown in Fig. \ref{fig_dataset_size}(b).}

\begin{figure}[!h]
\centerline{\includegraphics[width=0.7\columnwidth]{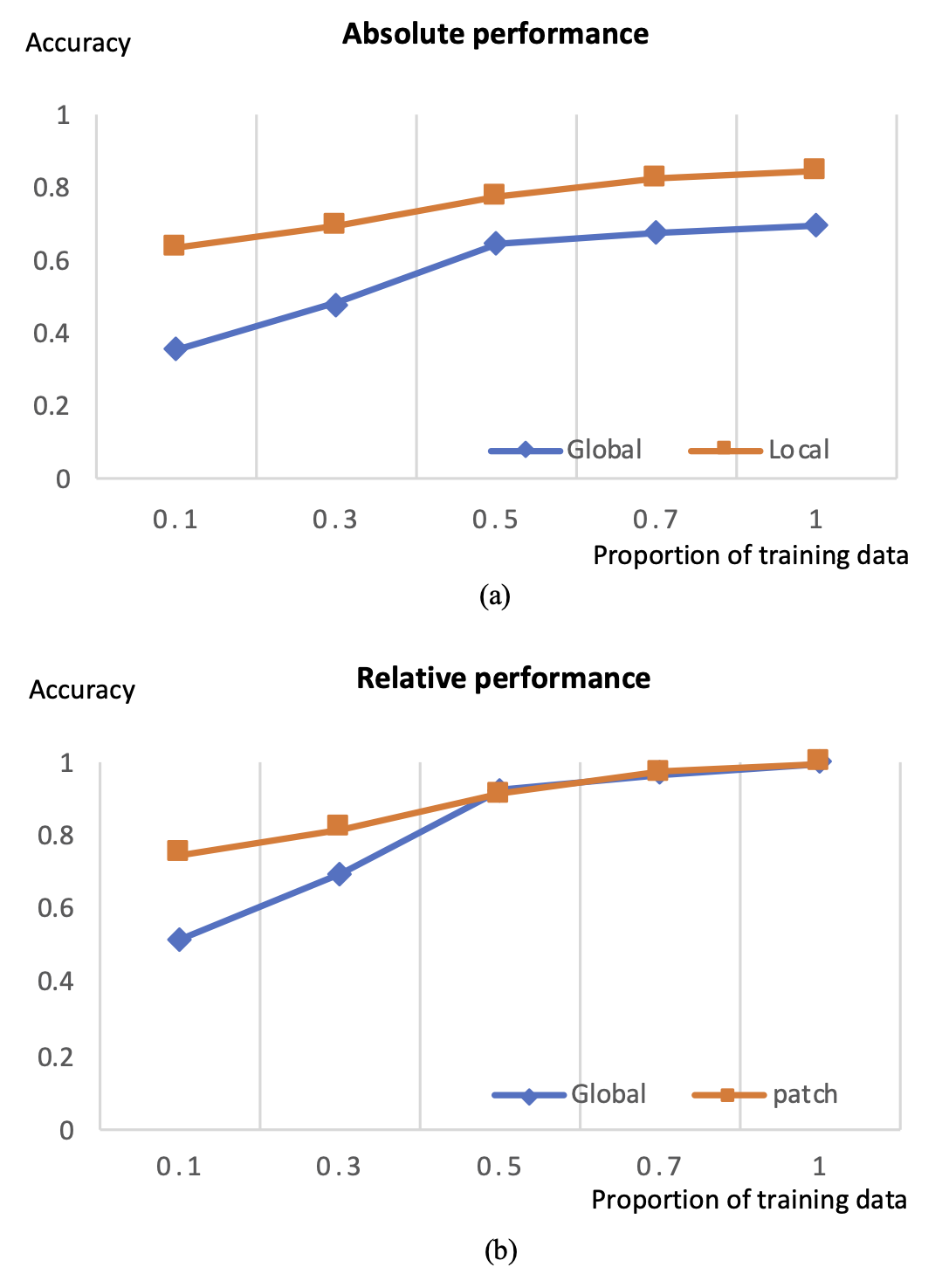}}
\caption{Accuracy of the algorithm according to the dataset size in (a)  absolute  and (b) relative scales.}
\label{fig_dataset_size}
\end{figure}

\section{Conclusion}

In the rapidly evolving global pandemic of COVID-19, the use of CXR for COVID-19 diagnosis or triage for patient management 
has become an important issue to preserve limited medical resources and prevent further spreading of the virus.
However, the current diagnostic performance with CXR is not sufficient for routine clinical use, so the need of artificial intelligence
to improve diagnostic performance of CXR is increasing.
Unfortunately, due to the emergent nature of COVID-19 global pandemic, systematic collection of the large data set for deep neural network training is difficult.  

To address this problem, we investigated potential biomarkers in the CXR and found  the globally distributed localized
intensity variation can be an discrimatory feature for the COVID-19. Based on this finding, we
propose a patch-based deep neural network architecture that can be stably trained with small data set. Once the neural network was trained, the final decision was made based on the majority voting from multiple patches at random locations within lungs. 
 {We also proposed a novel probabilistic Grad-CAM saliency map that is tailored to the local patch-based approach.}
Our experimental results demonstrated that the proposed network was trained stably with small data set,  provided comparative results with the SOTA method, and generated interpretable saliency maps that are strongly correlated with the radiological findings.

\appendices

\section*{Acknowledgment}
We would like to thank the  doctors and medical professionals from Korea and all around the world  who have dedicated their time and efforts to
treat COVID-19 patients and protect the health of citizens during this pandemic. 
This work was supported by the National Research Foundation
of Korea under Grant NRF-2020R1A2B5B03001980.



\bibliographystyle{IEEEtran}
\bibliography{COVID-19_bib}


\end{document}